\documentclass[preprint,journal]{vgtc}       % preprint (journal style)

\ifpdf%                                % if we use pdflatex
\pdfoutput=1\relax                   % create PDFs from pdfLaTeX
\usepackage{graphicx}                % allow us to embed graphics files
\DeclareGraphicsExtensions{.pdf,.png,.jpg,.jpeg} % for pdflatex we expect .pdf, .png, or .jpg files
\else%                                 % else we use pure latex
\usepackage{graphicx}                % allow us to embed graphics files
\DeclareGraphicsExtensions{.eps}     % for pure latex we expect eps files
\fi%

\graphicspath{{figures/}{pictures/}{images/}{./}} % where to search for the images

\usepackage{microtype}                 % use micro-typography (slightly more compact, better to read)
\PassOptionsToPackage{warn}{textcomp}  % to address font issues with \textrightarrow
\usepackage{textcomp}                  % use better special symbols
\usepackage{mathptmx}                  % use matching math font
\usepackage{times}                     % we use Times as the main font
\usepackage{cite}                      % needed to automatically sort the references
\usepackage{tabu}                      % only used for the table example
\usepackage{booktabs}                  % only used for the table example
\usepackage{graphics} % for pdf, bitmapped graphics files
\usepackage{epsfig} % for postscript graphics files
\usepackage{mathptmx} % assumes new font selection scheme installed
\usepackage{times} % assumes new font selection scheme installed
\usepackage{amsmath} % assumes amsmath package installed
\usepackage{amssymb}  % assumes amsmath package installed

\usepackage{microtype}                 % use micro-typography (slightly more compact, better to read)
\PassOptionsToPackage{warn}{textcomp}  % to address font issues with \textrightarrow
\usepackage{textcomp}                  % use better special symbols
\usepackage{mathptmx}                  % use matching math font
\usepackage{times}                     % we use Times as the main font
\usepackage{cite}                      % needed to automatically sort the references
\usepackage{tabu}                      % only used for the table example
\usepackage{amsmath}
\usepackage{booktabs}                  % only used for the table example
\usepackage{multirow}
\usepackage{color}
\usepackage[noend]{algpseudocode}

\usepackage{algorithmicx,algorithm}
\usepackage{balance}

\usepackage{enumitem}

\usepackage{soul}

\definecolor{revised}{RGB}{255,0,0}

\setenumerate[1]{itemsep=0pt,partopsep=0pt,parsep=\parskip,topsep=2pt}
\setitemize[1]{itemsep=0pt,partopsep=0pt,parsep=\parskip,topsep=2pt}
\setdescription{itemsep=0pt,partopsep=0pt,parsep=\parskip,topsep=2pt}
\onlineid{1180}

\vgtccategory{Research}
\vgtcpapertype{system}

\title{Strolling in Room-Scale VR: Hex-Core-MK1 Omnidirectional Treadmill}

\author{Ziyao Wang, Chiyi Liu, Jialiang Chen, Yao Yao, Dazheng Fang, Zhiyi Shi, \\Rui Yan, Yiye Wang, KanJian Zhang, Hai Wang, Haikun Wei*}
\authorfooter{%\vspace{-0.1cm}
\item
 Ziyao Wang, Chiyi Liu, Jialiang Chen, Yao Yao, Dazheng Fang, Zhiyi Shi, Rui Yan, Yiye Wang, KanJian Zhang and Haikun Wei are at Southeast University, China. E-mail: $\{$zy\_wang$|$cy\_liu$|$chenjialiang$|$yaoyao$|$\\fangdazheng$|$shizhiyi$|$yan\_rui$|$wangyiye$|$kjzhang$|$hkwei$\}$@seu.edu.cn
\item
 Hai Wang is at Saint Mary's University, Canada, E-mail: hwang@smu.ca.
\item
 Haikui Wei is the corresponding author.
 \vspace{-0.1cm}
}

\abstract{
	The natural locomotion interface is critical to the development of many VR applications. For household VR applications, there are two basic requirements: natural immersive experience and minimized space occupation. The existing locomotion strategies generally do not simultaneously satisfy these two requirements well. This paper presents a novel omnidirectional treadmill (ODT) system, named Hex-Core-MK1 (HCMK1).  By implementing two kinds of mirror symmetrical spiral rollers to generate the omnidirectional velocity field, this proposed system is capable of providing real walking experiences with a full-degree of freedom in an area as small as 1.76 m$\bf^2$, while delivering great advantages over several existing ODT systems in terms of weight, volume, latency and dynamic performance. Compared with the sizes of Infinadeck and HCP, the two best motor-driven ODTs so far, the 8 cm height of HCMK1 is only 20\% of Infinadeck and 50\% of HCP.  In addition, HCMK1 is a lightweight device weighing only 110 kg, which provides possibilities of further expanding VR scenarios, such as terrain simulation. The latency of HCMK1 is only 23ms. The experiments show that HCMK1 can deliver on a starting acceleration of 16.00 m/s$\bf^2$ and a braking acceleration of 30.00 m/s$\bf^2$. 
} % end of abstract


\teaser{
  \centering
  \includegraphics[width=\linewidth]{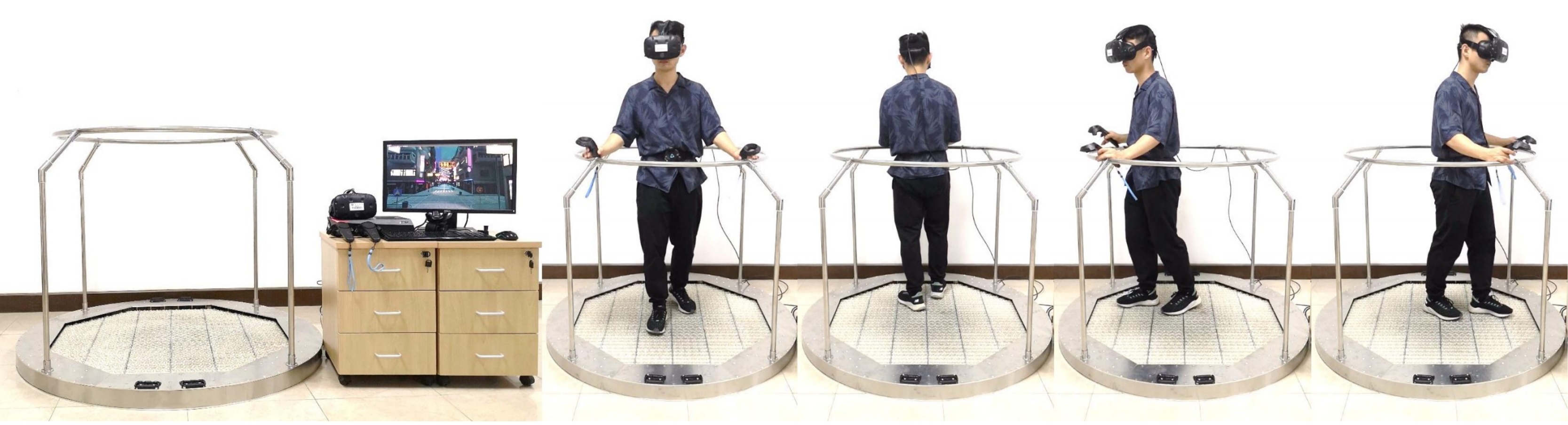}
  \caption{The left picture shows the complete Hex-Core-MK1 system and the right 4 pictures show the working status of the user walking along different directions. HCMK1 allows user to explore the infinite virtual world naturally in a small room without space constraints.}
	\label{fig:teaser}
 \vspace{-0.2cm}
}

\vgtcinsertpkg

\begin{document}

%% The ``\maketitle'' command must be the first command after the
%% ``\begin{document}'' command. It prepares and prints the title block.

%% the only exception to this rule is the \firstsection command

%\firstsection{Introduction}

\maketitle

\section{Introduction}	
%
%The rapid development of virtual reality (VR) has brought consumers an immersive experience. With the development of the head-mounted displays (HMDs) with higher resolution and higher refresh rate, users can see more realistic scenes. With the application of hand tracking, users can interact with the virtual environments (VE) using their own hands. VR technology has been applied in many fields, such as training, planning, education, psychology, sociology and entertainment \cite{Slater2016}. 
%However, current VR technology is still unnatural and not immersive enough to replace the real sense in daily life. One of the most important problems is the natural locomotion interface in VE.\cite{usoh1999walking}. 

With the development of virtual reality technology in recent years, the science fiction world in movies has gradually become a reality \cite{Slater2016}.In 2016, the virtual reality (VR) industry enjoyed a big explosion; then in 2017 and 2018, however, the industry’s bubbles went into burst, mainly due to such reasons as high prices of equipment, dearth of content and imperfect user experience (UX) \cite{Aric19,Kevin18,Benjamin2021}. 
The natural locomotion interface (NLI)\cite{usoh1999walking,steinicke2013human,hollerbach2002locomotion,iwata2000locomotion} is one of the subdivisions about the user experience in VR. Currently, limited by natural locomotion, most of the VR scenes are restricted to a fixed location or a small-scale local area, also compromising the richness of content. Problems with NLI have become a bottleneck for the development of many VR applications. Household VR applications often have to achieve natural locomotion in VR within a small physical space when trying to completely immerse users in the VR world. Deploying a small NLI solution with an immersive locomotion experience is crucial for further advancing the VR industry.
%\begin{figure}[tb]
%	\centering % avoid the use of \begin{center}...\end{center} and use \centering instead (more compact)
%	\includegraphics[width=\columnwidth]{pic/render.pdf}
%	\caption{Different color renderings of the HCMK1 omnidirectional treadmill. This is a circular ODT and the active area is a regular octagon. Compared with the square shape, this reduces the area waste on the diagonal. }
%	\label{fig:render}
%\end{figure}

The NLI problem has been extensively studied. Several strategies based on different principles have been proposed, such as the controller-based methods, the walking in large space methods, the walking in place methods, the redirected walking methods and the omnidirectional treadmill(ODT) methods \cite{al2018virtual}. Although these strategies can provide locomotion in VR, most cannot guarantee a reasonably good UX. The main limitations include motion sickness, low degree of freedom, high latency, and occupies a lot of space. %large space occupation.

For example, one simple controller-based locomotion method uses the controller, such as the touchpad, to control the movement in the virtual environment (VE), just like the classic control method ``WASD''\footnote{https://en.wikipedia.org/wiki/WASD}  or joystick in computer games. It can lead to a serious 3D motion sickness \cite{kolasinski1995simulator,langbehn2018evaluation,laviola2000discussion} due to the conflict between visual perception and inner ear perception
%since the eyes tell the brain that it senses the motion, but the inner ear transmits to the brain that everything is still. As a result of this incongruity, 
%the brain concludes that the individual is hallucinating and further concludes that the hallucination is due to poison ingestion. To clear the supposed toxin 
%the brain induces vomit 
\cite{Lawson2014a}. Another controller-based method is teleportation \cite{langbehn2018evaluation}. The limitation includes the lack of awareness on the intermediate path, which will affect the body’s positioning of the current location and led to getting lost.
Besides, the user may have to repeat additional operations to arrive at a proper position when they are reaching an object.

The Walk-in-place (WiP) methods replace natural walking with some specified actions, such as arm swinging \cite{wilson2016vr,pai2017armswing}, jogging\cite{hanson2019improving,kitson2017comparing,lee2018walking,nilsson2018natural}, etc. Since the WiP methods have the process of simulating walking, their UX is better than that of the handheld controller-based methods \cite{tregillus2016vr,hanson2019improving,wilson2016vr}. Generally, each proxy action can represent only one orientation movement. Basic movements like forward, backward and lateral movements (right and left) \cite{wang2018step} will need 4 proxy actions. 
Different orientation movements usually take the current view direction as a reference to apply different proxy actions\cite{templeman2006immersive}. Hence, it’s difficult for users to move in a straight line when looking around.
%Different orientation movements usually depend on the current view direction\cite{templeman2006immersive}, therefore the user is hard to keep moving along the original direction when changing their view direction. 
In addition, the starting and stopping latency is a challenge for the WiP methods\cite{templeman1999virtual,feasel2008llcm}, i.e., how long it takes the VR locomotion to start or stop after the user takes a step or stops their steps, or, to put it simply, the latency between the VR scene and the user's actual intended locomotion.
Latency in WiP is the result of adding a low-frequency filter or increasing the detection threshold in the data process, which aims to improve the detection accuracy and reduce the false triggers\cite{feasel2008llcm,wendt2010gud}. For example, the first proposed WiP method\cite{slater1994steps} does not start the VR locomotion until 4 consecutive steps had been taken. VR-STEP\cite{tregillus2016vr} averages every 5 samples to filter the signal, and the latency is about 100$ms$ to 200$ms$. LLCM-WIP\cite{feasel2008llcm} chooses a cut-off frequency of 5Hz, which adds about 100$ms$ of latency to both starting and stopping. After applying offsetting operation, the final starting latency reaches 138$ms$ and the stopping latency of 96$ms$. Generally, the WiP methods seek a compromise between 
false triggers and latency which are inevitable and difficult to eliminate.
%false triggers and latency, therefore, such latency usually is inevitable and difficult to eliminate.

Moving in VE with our own legs is the most natural way of locomotion interfaces\cite{iwata2000locomotion,hollerbach2002locomotion,steinicke2013human}. Based on the low latency tracking technology, walking in large space \cite{waller2007hive} is an intuitive solution. Generally, the navigation space in VE is equal to the physical space. The limitation of this method is that the real-world space will limit the virtual world space. If users moved close to the boundary of the physical space, they could not take more steps. In order to make full use of the limited space, the space compression technology, such as the redirected walking (RDW) approach, has been extensively studied in recent years. Based on the human body’s insensitivity to slight rotation and translation, the RDW approach can leverage visual dominance to subtly manipulate the user’s physical path\cite{bachmann2019multi,bolling2019shrinking}. When the user walks in a straight line in VE, the RDW approach guides the user to walk on a circle in the physical space. This is an effective strategy to compress the limitless virtual space to a limited physical space\cite{lee2019real,schmitz2018you,nilsson2018natural}. Generally, the RDW approach uses unperceived curvature gains to manipulate the user's path\cite{steinicke2009estimation}. It has been shown that the minimum demand for space is about 7$m$ in radius or 200$m^2$ \cite{bolling2019shrinking}. To further reduce the space required, Telewalk \cite{Rietzler2020} uses perceivable gains deliberately which highly reduces the space requirement to 3$m$$\times$3$m$. Since the Telewalk approach uses very high RDW gains which can lead to motion sickness symptoms \cite{Rietzler2020}, it sacrifices UX in exchange for the space reduction.

Strategies with auxiliary equipment usually have a much better performance in terms of the UX and space requirements. Omnidirectional treadmills (ODT) are a representative type of auxiliary equipment. The main idea is to keep the users' bodies stationary when they are walking \cite{al2018virtual}. Comparing with the walking in large space and RDW approaches, ODT could provide a similar walking experience, but requires much less space, and does not have any boundary restrictions. Comparing with the WiP methods, ODT occupies a slightly larger area but eliminates the limitations of the WiP strategy. For example, the VR locomotion of ODT is independent of the view direction, and the user can move in any direction without constraints. Hence, ODT achieves a higher degree of freedom than the WiP methods. In addition, ODT usually doesn't need to consider the false trigger problem associated with the WiP methods, so the starting and stopping latency is much lower. In general, as an auxiliary device, the ODT is capable of simultaneously ensuring the UX and space occupation, and it has been demonstrated to be an effective solution to the NLI problem in VR \cite{al2018virtual}.

Currently, a lot of different ODT design schemes have been proposed, and they can preliminarily achieve the basic functions. However, these design schemes have different limitations, such as laborious to use, dead zone, low degree of freedom, and bulkiness. These limitations are inherent shortcomings of the design and are difficult to eliminate through optimization of the manufacturing process.

To overcome these aforementioned limitations, this paper proposes a novel ODT design scheme, i.e., the 45-degree wheel-based scheme with the spiral rollers as the carrier, which has significant advantages in 
%solving the issues above especially 
terms of
the UX and volume. Based on this design scheme, we create the Hex-Core-MK1 (HCMK1) system. On the premise of providing the most natural locomotion experience, HCMK1 is a very small device that only occupies 1.76$m^2$ area and 8.0$cm$ height. The small volume indicates that it is suitable for the room-scale VR. The 110$kg$ weight ensures high dynamic performance and provides capabilities for many VR applications like terrain simulations. In addition, the starting and stopping latency of HCMK1 is only 23$ms$. Low latency ensures the VR scenes can be consistent with the user's actual intention in real-time. This paper further also analyzes the main factors affecting the UX, includes the OVF(omnidirectional velocity field) working delay and the height of the user's center of gravity, which may guide the design of the controller to further improve the UX.

The main contributions of this paper are as follows:
\begin{itemize}
	%\vspace{-0.03cm}
	\setlength{\topsep}{0pt}
	\setlength{\itemsep}{0pt}
	\setlength{\parsep}{0pt}
	\setlength{\parskip}{0pt}
	\item Proposing a novel design scheme of driven based ODT. 
	\item Developing the HCMK1 system based on the proposed scheme, which has the state of the art performance in terms of volume, weight, latency and dynamic performance.
	\item Analyzing the main factors affecting the UX based on the survey and the experimental data. 
	%\vspace{-0.16cm}
\end{itemize}

The remainder of the paper is organized as follows.  Section 2 describes the related work.  Section 3 presents the proposed novel design scheme of the HCMK1 system. Section 4 compares the proposed HCMK1 system with other driven-based ODTs.  Section 5 presents an experimental study, and Section 6 presents the limitation of the proposed HCMK1 system, respectively.  Finally, Section 7 summarizes the results and highlights the future work. In addition, several videos about an application demo and part of the experiments are attached in the supplementary material.

\begin{figure*}[htb]
	\centering % avoid the use of \begin{center}...\end{center} and use \centering instead (more compact)
	\includegraphics[width=\linewidth]{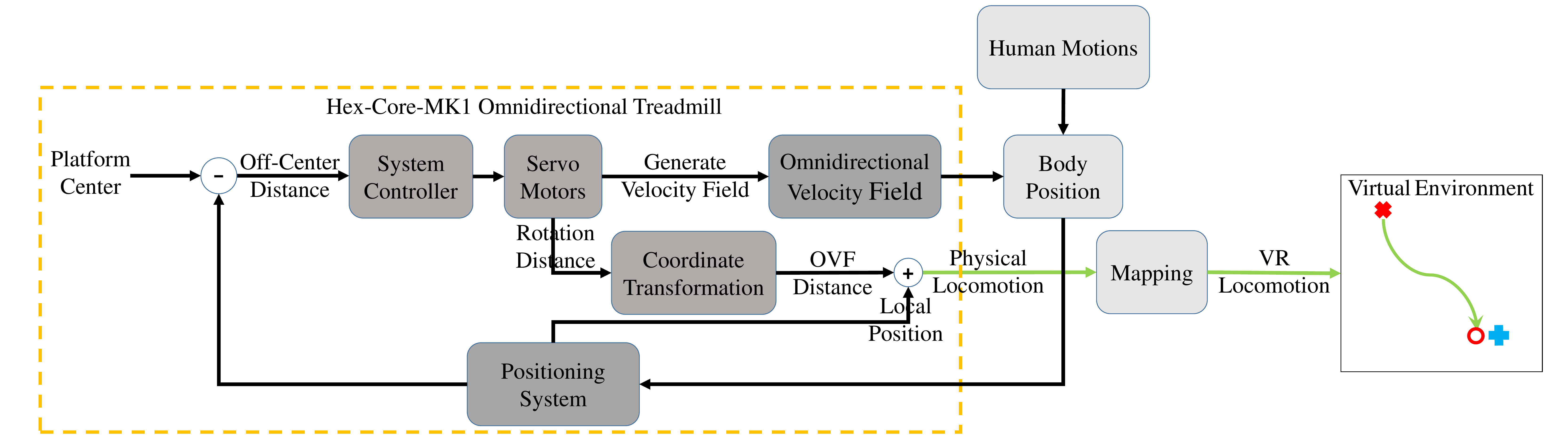}
	%\vspace{-0.6cm}
	\caption{The flow chart of HCMK1 system. The orange dash line shows the complete working process, which interacts with human. The external disturbance is the changes in the body’s position caused by motions. HCMK1 aims to reduce this disturbance by carrying the user back to the center. The positioning system detects the body’s position to calculate the current off-center distance. Based on the off-center distance, the system controller controls the servo motors’ rotation speed. The special mechanical structure further converts the parallel rotation speed into OVF, which finally provides the driving force to carry the user back to the center in any direction. The rotation distance of servo motors reflects the distance of OVF provides. The physical locomotion distance of the user could be calculated by the OVF distance and their local position. When applying the physical locomotion distance to the VE, a mapping process is necessary, which mainly transforms the coordinate, adjusts the locomotion gains, or introduces some nonlinear functions to improve the UX.		
		%		Firstly, the positioning system detects this position changes. Then based on the current position, the controller calculates the off-center distance of the user and adjust the servo motors' rotation speed. The omnidirectional velocity field composites the rotation speed of servo motors to a single velocity with a certain direction and amplitude based on its special mechanical structure. This velocity carries the user back to the center. The output data of HCMK1 is the calculated physical locomotion distance of the user. The mapping process between the physical locomotion and the VR locomotion is determined by the VR scene. Different mapping functions could achieve different experiences. Therefore, the mapping process is set outside the HCMK1 working process. 
	}
	\label{fig:systemflow}
	\vspace{-0.35cm}
\end{figure*}

\section{Related Work}
%\vspace{-0.08cm}
The main feature of ODT is to keep the user stationary when the user is walking. A simple and crude method is slide in place, i.e., the low-friction surface scheme \cite{huang2003omnidirectional,cakmak2014cyberith,ang2018put}. Generally, the user's waist is tied to the machine. The user needs to wear a pair of roller shoes or stand on a special surface with suitable low-friction, such as the polytetrafluoroethylene (PTFE) surface, the ball-bearing surface. When walking forward, the user’s feet will slide on the surface with the assistance of their waist. The commercial products based on this principle include Kat Walk, Virtuix Omni, Cyberith Virtualizer, etc. The main problem of this strategy is that overcoming the friction is too laborious, imagine the user needs to moonwalk on the surface like Jackson all the time. Users can slide forward laboriously but too hard to slide backwards or sideways \cite{cakmak2014cyberith}. The walking tutorial of Kat Walk \footnote{https://download.katvr.com/product/literature/KAT Walk mini Walking Tutorials1575006763475.mp4} shows the user needs to combine the WiP proxy actions to realize the backward and sideways. Besides, since the gait action is abnormal, it is hard to track the gait distance accurately. Kat Walk \footnote{https://www.manualslib.com/manual/1474138/Kat-Vr-Walk.html} and Virtuix Omni \footnote{https://www.roadtovr.com/virtuix-omni-preview-production-model-video/} use the inertial measurement unit (IMU) to count the number of steps and indirectly calculate the movement distance, which further reduces the accuracy of gait matching or even appears the reverse direction gaits.

Like the traditional treadmill in the gym, if ODT can automatically carry users back to the center, the UX will be much better. This kind of ODTs also called driven based ODT. CirculaFloor  \cite{iwata2005circulafloor} uses 4 independent autonomous circulating robots to carry the user back. The user always stands on 2 robots and another 2 robots move to the position of the user’s next step. String Walker \cite{iwata2007string} uses several strings to pull the shoes back to the center. CyberCarpet  \cite{schwaiger20072d,de2012motion} and StriderVR are two similar ODTs. They all lay a layer of steel balls on a traditional treadmill. The main difference is that when the user turns, CyberCarpet rotates the traditional treadmill in the same direction to provide velocity in a proper direction, but StriderVR rotates the steel-ball layer to the inverse direction to eliminate the user's rotation. The main problems of these ODTs are the low dynamic performance and the low degree of freedom (DOF).

The fast-responding and continuous omnidirectional velocity field (OVF) is the key to driven-based ODTs \cite{Wang2020}. Omnideck \cite{brinks2016redesign} can generate a fixed inwardly contracting OVF based on a number of inwardly rotating rollers. The fixed OVF causes the user can only walk on the outer ring of the surface, which leads to low area utilization and large area occupation. An ideal OVF should be parallel, continuous, and able to quickly respond to direction changes. A traditional scheme is the belt-based ODT \cite{darken1997omni}, which can be simply understood as a big treadmill (x-axis) carrying several small treadmills (y-axis). This scheme has been applied in several previous works, such as the  Torus\cite{iwata1999walking}, Cyberwalk \cite{souman2011cyberwalk,schwaiger2007cyberwalk} and F-ODT \cite{lee2016design,pyo2018development}. The commercial product Infinadeck \cite{metsis2017integration} is also based on this scheme. This scheme could provide a real walking experience with full DOF. The only drawback is that it is hard to miniaturize due to its double-layer structure. As a commercial product, Infinadeck has been optimized for several years and still has 40$cm$ height and 225$kg$ weight. The large volume limits its application areas. Another scheme is the 45-degree wheel-based scheme \cite{Wang2020}. Hex-Core-Prototype (HCP) is a much thinner and smaller ODT based on this scheme. The main components of HCP are the mirror-symmetrical chain and the small wheels arranged at 45 degrees on the chain. It can generate the ideal OVF based on the principle of the decomposition and composition of the velocity. HCP has reduced the height to 16$cm$ and the weight to 150$kg$, which proves that the 45-degree wheel-based scheme has more competitive than the belt-based scheme in terms of miniaturization. However, the chain usually stretches soon which leads to the tooth-jumping problem. Taking the chains as carrier still is a double-layer structure, therefore, the volume could be further reduced.

%
%

%
%This paper first presents the design of the HCMK1 system, including the design of an omnidirectional velocity field, positioning system, a basic controller, locomotion calculation process and the latency analysis. Then we present the developed system and compare HCMK1 with several other driven-based ODTs. This part demonstrates HCMK1 has great advantages in weight, volume, dynamic performance and low latency. The next section uses several experiments to test this system and further analyzes the main factors affecting the UX based on the recorded data. In addition, we develop an application demo to demonstrate the complete functions of the HCMK1 system. The limitations of the current system are demonstrated in the next section. The last section mainly discusses the work that needs to improve and the outlook for future work.
%

%This paper first presents the novel design scheme of the HCMK1 system. Then we present the developed system and compare HCMK1 with several other driven-based ODTs. The next section includes several experiments about the HCMK1 basic function and UX. Then this paper lists the limitations of the current version. The last section mainly discusses the work that needs to improve and the outlook for future work. In addition, several videos about an application demo and part of experiments are attached in the supplementary material. 
%\vspace{-0.1cm}

\begin{figure*}[ht]
	\centering % avoid the use of \begin{center}...\end{center} and use \centering instead (more compact)
	\includegraphics[width=\linewidth]{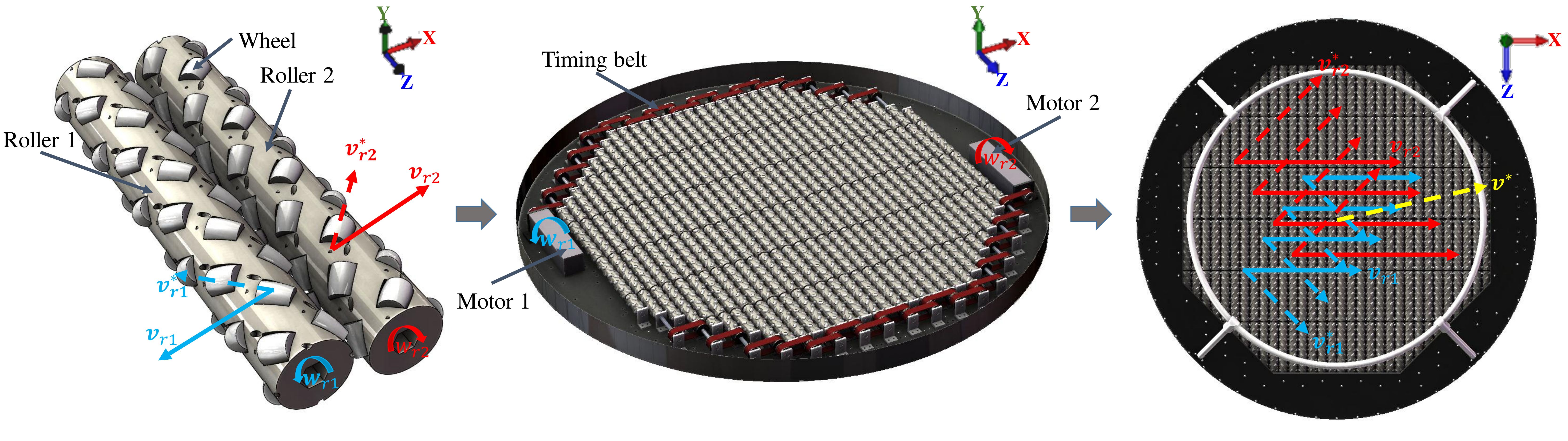}
	%\vspace{-0.9cm}
	\caption{The mechanical design of the omnidirectional velocity field. The left image demonstrates the core components of our scheme, i.e., the  mirror-symmetrical spiral rollers. The roller body is embedded with a number of small wheels which are arranged at an angle of $\pm$ 45 degrees. This is a compact structure and the radius is only 1.71$cm$. The mid image shows that these two kinds of rollers are driven by two identical servo motors respectively. The right figure shows the system surface structure after arranging the two kinds of rollers alternately and demonstrates the composition process of the OVF velocity. }
	\label{fig:OVF}
	%\vspace{-0.6cm}
\end{figure*}

\section{System Design}

HCMK1 is also based on the 45-degree wheel-based scheme. The main improvement is on the carriers of the 45-degree wheels. We design a pair of spiral rollers to support the 45-degree wheels. Compared with the mirror-symmetrical chains, the spiral rollers have a more compact structure and more stable performance. It reduces the weight of the rotation components and greatly reduces the moment of inertia through a small rotation radius, only 1.71$cm$. Therefore, HCMK1 has a much better dynamic performance.

%\vspace{-0.1cm}
Figure \ref{fig:systemflow} presents the workflow of HCMK1. As a closed-loop control system, HCMK1 consists of 4 main parts, i.e., the omnidirectional velocity field, the positioning system, the system controller and the locomotion calculation process. This section discusses these 4 parts as well as the analysis the latency.
% The positioning system detects the body position to calculate the current off-center distance. Based on the off-center distance, the controller controls the servo motors to generate the velocity field. The special mechanical structure further converts the parallel velocity field into OVF, which finally  provides the driving force to carry the user back to the center in any direction. The rotation distance of servo motors reflects the distance of OVF provides. The physical locomotion distance of the user could be calculated by the OVF distance and their local position. When applying the physical locomotion distance to the VE, a mapping process is necessary, which mainly transforms the coordinate, adjusts the locomotion gains, or introduces some nonlinear functions to improve the UX.
%\vspace{-0.1cm}
\subsection{Notations}
In this paper, we use $x,\textbf{x}$ to denote the scalar and vector values respectively. %The symbol represents its physical property.
The subscript of a notation indicates the source of the symbol, which is usually the first letter or abbreviation. Table \ref{tab:notation} summarizes the notations used in this paper.

%
%$\mathbf{w}_{\text{r1}}$,
%$\mathbf{v}_{\text{r1}}$,
%$\mathbf{v}^*_{\text{r1}}$ denote the angular velocity, the surface linear velocity and the retained velocity that along the wheel’s axis of Roller1.

%$\theta_{\text{r1}}$ denotes the angle of the wheels on Roller1.

%$\mathbf{w}_{\text{r2}}$,
%$\mathbf{v}_{\text{r2}}$,
%$\mathbf{v}^*_{\text{r2}}$,$\theta_{\text{r1}}$ denote the same meanings but of Roller2. 

%$d_\text{r}$ denotes the diameter of Roller1 and Roller2.

%$\theta_{\text{r1}},\theta_{\text{r2}}$

%$\mathbf{w}_{\text{r1}}=(0,0,w_{r1})$ 
%
%$\mathbf{w}_{\text{r2}}=(0,0,w_{r2})$

%$\mathbf{v}^{*}_{\text{ovf}}$ denotes the velocity of OVF with the direction of $\theta^{*}$ and the amplitude of 
%$\alpha$.

%$-------------------------------$

%$X_{tr}Y_{tr}Z_{tr}$
%
%$\mathbf{p}_{\text{utrl}}$
%
%$\mathbf{p}_{\text{tg}}$
%
%$\mathbf{q}_{\text{tg}}$
%
%$\mathbf{p}_{\text{ug}}$

%$-------------------------------$
%
%$\mathbf{p}_{\text{off}}$
%
%$\mathbf{p}_{\text{ref}}$
%
%$D_{th}$ 
%
%$K_p=2$ 

%$-------------------------------$
%
%$n_1\left(t\right)$ 
%
%$n_2\left(t\right)$

%$\mathbf{D}^*_{\text{ovf}}(t)$

%$\lambda$

%$\mathbf{D}_{\text{pe}}(t)$
%
%$\mathbf{D}_{\text{ve}}(t)$
%
%$\mathbf{D}_{\text{ovf}}(t)$
%
%$\mathbf{D}_{\text{rm}}(t)$

%$\mathbf{S}_L(t)\in R^{3\times3}$

%$\mathbf{v}^{*}_{\text{ovf}}(t)$
%
%$\int_{0}^{t}\mathbf{v}^{*}_{\text{ovf}}(x)dx=\mathbf{D}^*_{\text{ovf}}(t)$

%$-------------------------------$

%$T_\text{m}=\max(T_{\text{rm}},T_{\text{ovf}})=22ms$
%
%$T_{\text{c}}=1ms$
%
%$T_{\text{s}}=23ms$
%
%$-------------------------------$

%$\beta_{1} < \beta_{2}$

%\vspace{-0.15cm}

\subsection{Design of Omnidirectional Velocity Field}
%\vspace{-0.1cm}
The advantages of HCMK1 in terms of the UX and volume depend on the design of OVF. The left part of Figure \ref{fig:OVF} shows the main components, that is the mirror-symmetrical spiral rollers. The roller body is embedded with a number of small wheels, which are arranged at an angle of $\pm$ 45 degrees. Compared with the mirror-symmetrical chain in the HCP system \cite{Wang2020}, this is a more compact structure with a radius of only 1.71$cm$. Therefore, the overall volume and weight of the HCMK1 system can be further reduced.

\vspace{0.05cm}
\begin{table}[htp]
	\caption{Notations in this paper.}
	\scriptsize%
	\label{tab:notation}
	\centering  
	%\vspace{-0.1cm}
	\begin{tabular}{m{0.18\linewidth} m{0.74\linewidth}}
		\toprule
		\multicolumn{1}{c}{Notation} & \multicolumn{1}{c}{Meaning}  \\ \midrule
		$\mathbf{w}_{\text{r1}}$, $\mathbf{v}_{\text{r1}}$, $\mathbf{v}^*_{\text{r1}}$     &$\diamond$ Angular velocity, the surface linear velocity and the retained velocity that along the wheel’s axis of Roller1.                \\
		$\mathbf{w}_{\text{r2}}$, $\mathbf{v}_{\text{r2}}$, $\mathbf{v}^*_{\text{r2}}$ 	  &$\diamond$   The same meanings as above but of Roller2. \\
		$\theta_{\text{r1}}$, $\theta_{\text{r2}}$     							  &$\diamond$ Angle of the wheels on Roller1 and Roller2. \\
		$d_\text{r}$  												& $\diamond$ Diameter of Roller1 and Roller2.\\	      
		$\mathbf{v}^{*}_{\text{ovf}}$, $\theta^{*}$, $\alpha$             &  $\diamond$ Velocity of OVF with the direction of $\theta^{*}$ and the amplitude of $\alpha$. \\
		$\mathbf{p}_{\text{utrl}}$                                        &  $\diamond$ User's position in the tracker's coordinate system, i.e., $X_{tr}Y_{tr}Z_{tr}$ coordinate system.\\
		$\mathbf{p}_{\text{tg}}$,$\mathbf{q}_{\text{tg}}$ 	 					& $\diamond$ Tracker's position and spatial attitude in the global coordinate system. \\
		$\mathbf{p}_{\text{ug}}$  	&  $\diamond$ User's position in the global coordinate system. \\
		$\mathbf{p}_{\text{ref}}$ 	&  $\diamond$ A preset point on X-Z plane. User will be sent to this point. \\
		$\mathbf{p}_{\text{off}}$  	&  $\diamond$ The offset position between user's position and the preset reference point. \\
		$n_1\left(t\right)$,  $n_2\left(t\right)$  				&  $\diamond$ The number of revolutions of servo motor1 and motor2. \\
		$\lambda$  			&  $\diamond$ Reduction ratio of servo motors. \\
		$\mathbf{D}^*_{\text{ovf}}(t)$  &  $\diamond$ The locomotion distance provided by the OVF. \\
		$\mathbf{D}_{{\text{ovf}}}(t)$  &  $\diamond$ The user's locomotion distance that offset by the OVF. \\
		$\mathbf{D}_{\text{pe}}(t)$, $\mathbf{D}_{\text{ve}}(t)$  				&  $\diamond$ The user's locomotion distance in physical environment and virtual environment. \\
		$\mathbf{D}_{\text{rm}}(t)$  										&  $\diamond$ The user's local locomotion distance that has not been offset by the OVF.\\
		%		 $\mathbf{S}_L(t)$  										&  The sliding loss when the user walking on the OVF. \\
		$T_\text{m}$, $T_{\text{rm}}$, $T_{\text{ovf}}$  								&  $\diamond$ The latency of the total measurements, the positioning measurement, and the motor measurement. \\
		$T_{\text{c}}$ 					&   $\diamond$ The latency of serial communication. \\
		$T_{\text{s}}$  				&  $\diamond$ The latency of the system or the latency between the VR scenes and the user's intended locomotion.\\
		$\beta_{1}$, $\beta_{2}$  & $\diamond$ Gains of the platform OVF locomotion and the user's local locomotion. \\
		\bottomrule
	\end{tabular}
	%\vspace{-0.7cm}
\end{table}

To put it simply, OVF is composed of two types of rollers, i.e., Roller1 type and Roller2 type in the left image of Figure \ref{fig:OVF}, which are alternately densely arranged. 
All rollers of the same type are connected by timing belts and driven by the same motor, so they have the same speed. The main difference between these two types of rollers is the direction of the wheel embedded on the surface. All of the wheels are not powered and can rotate around their axes freely. Therefore, for the external speed, the wheel can counteract the speed component in the direction of rotation, and only the axial speed component is retained. This is similar to the principle of the Mecanum wheel \cite{ilon1975wheels}. After such a decomposition process, the remained speed components of these two types of rollers are perpendicular to each other. By adjusting the speed amplitude, the OVF speed in any direction is composited.

The following discussion is based on the right-handed coordinate system in the figure. When the Roller1 rotates around the z-axis at the angular velocity of $\mathbf{w}_{\text{r1}}$, the linear velocity on the surface, i.e., the x-z plane, is $\mathbf{v}_{\text{r1}}$. Since the 45-degree wheel rotates freely, $\mathbf{v}_{\text{r1}}$ can be decomposed into two mutually perpendicular velocities, where the velocity perpendicular to the wheel’s axis will be counteracted, and another velocity that along the wheel’s axis, i.e., the blue dash line, $\mathbf{v}^*_{\text{r1}}$, is retained. Similarly, when the Roller2 rotates at the angular velocity of $\mathbf{w}_{\text{r2}}$, the surface linear velocity is $\mathbf{v}_{\text{r2}}$, only $\mathbf{v}^*_{\text{r2}}$, i.e., the red dash line, is retained. Based on the spiral rollers, the originally two parallel velocities $\mathbf{v}_{\text{r1}}$ and $\mathbf{v}_{\text{r2}}$ are decomposed into two mutually perpendicular velocities $\mathbf{v}^*_{\text{r1}}$ and $\mathbf{v}^*_{\text{r2}}$, which could be calculated by
%\vspace{-0.35cm}
\begin{equation}
%\vspace{-0.2cm}
\begin{split}
&\mathbf{v}_{\text{r1}}=\mathbf{w}_{\text{r1}}\times(0,\frac{d_\text{r}}{2},0);    \mathbf{v}_{\text{r2}}=\mathbf{w}_{\text{r2}}\times(0,\frac{d_\text{r}}{2},0) ;\\
&\mathbf{v}^*_{\text{r1}}=\langle\left(\cos(\theta_{\text{r1}}),0,\sin(\theta_{\text{r1}})\right), \mathbf{v}_{\text{r1}}\rangle\times\left(\cos(\theta_{\text{r1}}),0,\sin(\theta_{\text{r1}})\right);\\
&\mathbf{v}^*_{\text{r2}}=\langle\left(\cos(\theta_{\text{r2}}),0,\sin(\theta_{\text{r2}})\right), \mathbf{v}_{\text{r2}}\rangle\times\left(\cos(\theta_{\text{r2}}),0,\sin(\theta_{\text{r2}})\right).
\end{split}
\label{eqn:verticalvelocities}
\end{equation}
Here $d_\text{r}=3.42cm$ denotes the diameter of the rollers, $\theta_{\text{r1}}=\frac{\pi}{4}$ and $\theta_{\text{r2}}=-\frac{\pi}{4}$ denote the angle of the wheels on Roller1 and Roller2. $\mathbf{w}_{\text{r1}}$ and $\mathbf{w}_{\text{r2}}$ denotes the angular velocities which point along the z-axis, the preset conditions are $\mathbf{w}_{\text{r1}}=(0,0,w_{r1})$ and $\mathbf{w}_{\text{r2}}=(0,0,w_{r2})$.

Arranging these two kinds of rollers alternately could construct a surface shown in the middle of Figure \ref{fig:OVF}. The same type rollers are connected in series using a row of timing belts and driven by one servo motor. Therefore, HCMK1 only needs two motors to construct the OVF, and the same type rollers have the same velocity. 
%We have attempted to use gear and chain as the transmission mechanism. The main problem is that the gear scheme requires high assembly accuracy, and a driven gear needs to be installed between every two gears to ensure the same direction of rotation. In the first generation of HCMK1, when the user stood on the platform, the gear often stuck. This led to the failure of the first generation of HCMK1. The chain scheme could solve the stuck problem, but the chains were stretched soon, and the tooth-jumping occurs. It makes a loud noise. Finally, we applied the timing belt, which is quieter when working.

The right part of Figure \ref{fig:OVF} presents the composition process of OVF. $\mathbf{v}^*_{\text{r1}}$ and $\mathbf{v}^*_{\text{r2}}$ are staggered on the plane. When these two velocities act on the same object and ignore the torque caused by the non-coincidence of the acting positions, it is easy to get the
composited velocity $\mathbf{v}^{*}_{\text{ovf}}=\mathbf{v}^*_{\text{r1}}+\mathbf{v}^*_{\text{r2}}$. Conversely, when a velocity with the direction of $\theta^{*}$ and an amplitude of $\alpha$ is required, i.e., $\mathbf{v}^{*}_{\text{ovf}}=(\alpha \cos(\theta^{*}),0,\alpha \sin(\theta^{*}))$, these two type spiral rollers need to be set at the angular velocities
%\vspace{-0.25cm}
\begin{equation}
%\vspace{-0.2cm}
\begin{split}
&\mathbf{w}_{\text{r1}}=(0,0,-\frac{2\alpha}{d_\text{r}}\cos(\theta^{*})-\frac{2\alpha}{d_\text{r}}\sin(\theta^{*})\tan (\theta_{\text{r1}}) );\\
&\mathbf{w}_{\text{r2}}=(0,0,-\frac{2\alpha}{d_\text{r}}\cos(\theta^{*})-\frac{2\alpha}{d_\text{r}}\sin(\theta^{*})\tan (\theta_{\text{r2}}) ).
\end{split}
\label{eqn:motorvelocityfromsynthesis}
\end{equation}
%
%In the HCMK1 system, the teeth number of the synchronous wheels in servo motor and spiral rollers is the same, hence Equation \ref{eqn:motorvelocityfromsynthesis} can be used to set the output angular velocities of the servo motor.

\begin{figure}[htb]
	\vspace{-0.2cm}
	\centering % avoid the use of \begin{center}...\end{center} and use \centering instead (more compact)
	\includegraphics[width=0.98\linewidth]{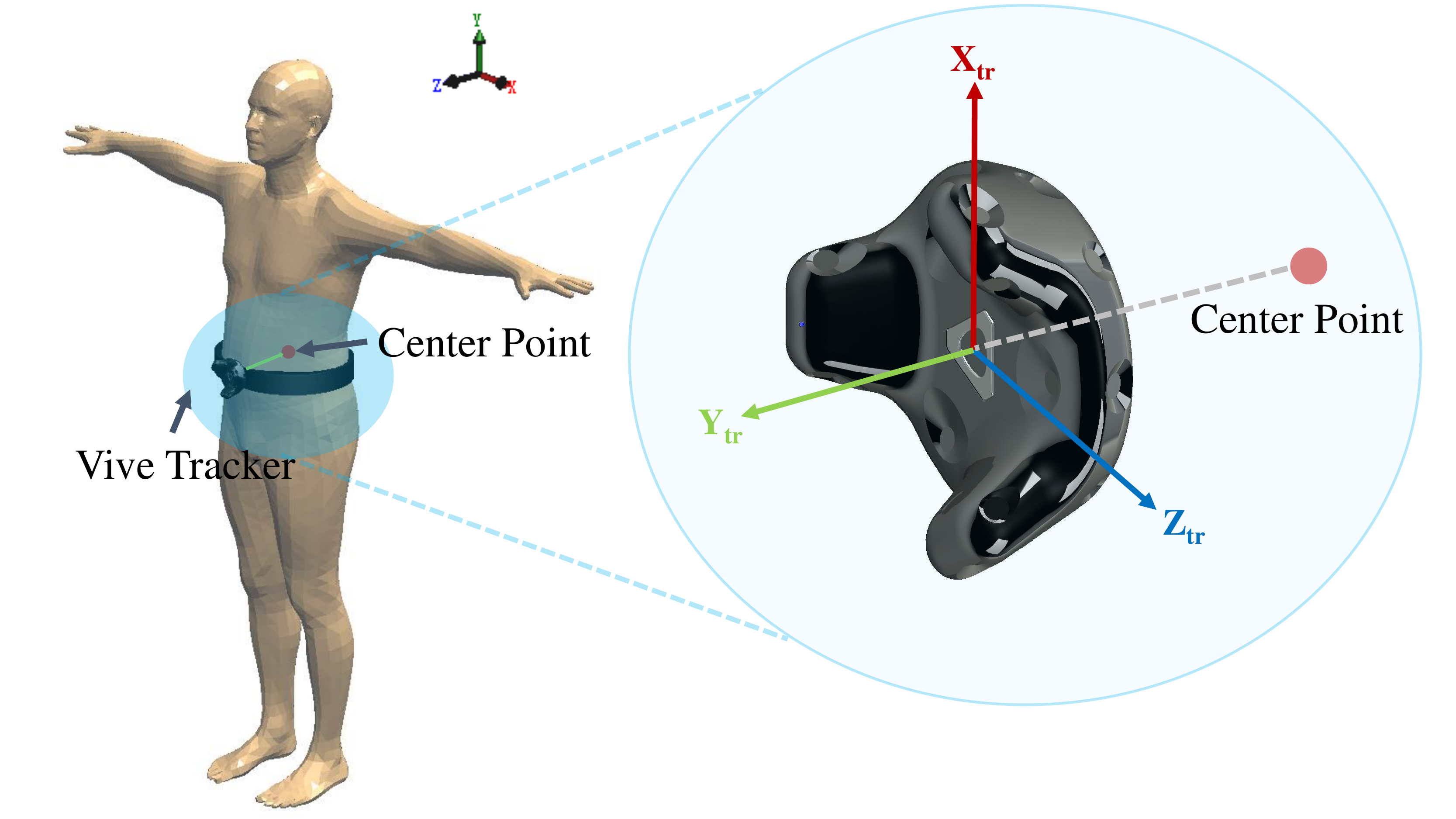}
	%\vspace{-0.45cm}
	\caption{The wearing method of vive tracker. In this way, the center point of the human body is a fixed point relative to the tracker coordinate system. 
		%		We can calculate the real-time global position based on the spatial position and attitude of the tracker. 
	}
	\label{fig:positioningsystem}
	\vspace{-0.1cm}
\end{figure}
\subsection{Positioning System}
%\vspace{-0.1cm}
High accuracy and low-latency positioning signals are important to the control performance. The HCMK1 system uses the Vive Tracker \footnote{https://www.vive.com/us/accessory/vive-tracker/} to track the user's position. It is based on the SteamVR tracking technology, which has a latency of less than 20 milliseconds and millimetre-scale positioning accuracy.

%As figure \ref{fig:positioningsystem} shows, we chose the center point to represent whole body. Because 
%the center of mass of the body determines the balance of human body. Besides, the active area of the platform is about 1.2 $m$ diameters, it is essential to take an accurate method to represent the user's position
%the active area of the platform is about 1.2 $m$ diameters, it is essential to take an accurate method to represent the user's position. Since the balance of human body is determined by the position of the center of mass. Therefore, measuring the position of the center of mass is the most accurate and convenient method. Though introduce the position information of the knees and feet could help to analyze the users' motion intention, but it is much more complicated. It could be taken as an improvement of the current positioning system in the future.

Since the active area of the platform is about 1.2$m$ in diameter, it is essential to take an accurate method to represent the user's position. As shown in figure \ref{fig:positioningsystem}, we chose the center point to represent the whole body. Because the balance of the human body is determined by the position of the center of the mass, measuring the position of the center portion is sufficient to represent the position of the human body. Though introduce the position information of the knees and feet could help to analyze the users' intention, but it is much more complicated, and it could be taken as an improvement of the positioning system in the future work.

The tracker is worn around the waist to track the center point's real-time position in the global coordinate system, i.e., in the $XYZ$-coordinate system. In this way of wearing, the center point is approximately fixed in the local coordinate system of the tracker, i.e., the $X_{tr}Y_{tr}Z_{tr}$-coordinate system. Record the local position of the center point in $X_{tr}Y_{tr}Z_{tr}$-coordinate system as $\mathbf{p}_{\text{utrl}}$, which is a constant. Record the global spatial position of the tracker as $\mathbf{p}_{\text{tg}}$ and use the quaternion to record the global spatial attitude as $\mathbf{q}_{\text{tg}}$. Here we chose the quaternion to avoid the gimbal lock of Euler angles and to get a more stable and more efficient calculation process than rotation matrices. Based on the arithmetic rules of quaternion, the global spatial position of the center point is denoted by
%\vspace{-0.25cm}
\begin{equation}
%\vspace{-0.25cm}
\begin{split}
&\mathbf{p}_{\text{ug}}=\mathbf{p}_{\text{tg}}+vector(\mathbf{q}_{\text{tg}}*Quaternion(0,\mathbf{p}_{\text{utrl}})*\mathbf{q}^{-1}_{tg}),\\
\end{split}
\label{eqn:positioning}
\end{equation}
where the $Quaternion(w,(x,y,z))$ means generating a quaternion, i.e., $w+xi+yj+zk$ and $vector(\cdot)$ means to extract the vector part of the quaternion. $\mathbf{p}_{\text{utrl}}$ is an approximate value, and preset as $\mathbf{p}_{\text{utrl}}=(0,-0.1,0)$.

%\vspace{-0.1cm}
\subsection{Basic Controller \label{Basic Controller}}
%\vspace{-0.1cm}
An ideal experience of ODTs is when the user abruptly starts walking or stops walking at a normal speed, the body always keep stationary. This is a complex human-computer interaction problem that has a high requirement for the dynamic performance of the platform and it also needs to combine the kinematic model of the human body. 

To preliminary verify this design scheme, we designed a basic controller. This is a simple proportional controller with offset. When the user is at the position $\mathbf{p}_{\text{ug}}$, set the OVF at the velocity
%\vspace{-0.3cm}
\begin{equation}
%\vspace{-0.2cm}
\begin{split}
&\mathbf{v}^{*}_{\text{ovf}}=\alpha\cdot\frac{\mathbf{p}_{\text{off}}}{{\|\mathbf{p}_{\text{off}}\|}_2}.\\
\end{split}
\label{eqn:v*}
\end{equation}
\begin{equation}
%\vspace{-0.1cm}
\begin{split}
&\mathbf{p}_{\text{off}}=\left[  \begin{array}{ccc}
1 & 0 & 0\\
0 & 0 & 0\\
0 & 0 & 1\\ \end{array}  \right]\cdot\left(\mathbf{p}_{\text{ref}}-\mathbf{p}_{\text{ug}}  \right),\\
\end{split}
\label{eqn:poff}
\end{equation}
\begin{equation}
%\vspace{-0.1cm}
\begin{split}
&\alpha=\left\{ \begin{array}{ccl}
0 & \mbox{for}& \|{\mathbf{p}_{\text{off}}\|}_2<D_{th} \\ 
K_p\cdot\left(\|{\mathbf{p}_{\text{off}}\|}_2-D_{th}\right) & \mbox{for} & \|{\mathbf{p}_{\text{off}}\|}_2\geq D_{th} \\
\end{array}\right. .\\
\end{split}
\label{eqn:alpha}
\end{equation}
Here $\|\cdot\|$ denotes the Euclidean norm of vectors, $\mathbf{p}_{\text{off}}$ denotes the offset position between the projection of the user's position on the X-Z plane and a preset reference point, i.e., $\mathbf{p}_{\text{ref}}$. In this basic controller, we set $\mathbf{p}_{\text{ref}}$ as the center of the OVF. $\alpha$ denotes the amplitude of the velocity. It is set as a piecewise function to avoid frequent adjustments when the user close to the reference point. 
%Because even though the user stands still in place, their body still has some movements in a small range. To avoid frequent adjustments, when the user stands close to the center of the OVF, the OVF should keep stationary. 
This is based on the control threshold $D_{th}$. According to the experiments, setting $D_{th}=0.08$ is sufficient to avoid frequent adjustments. Considering the safety factor, we set $K_p=2$, which denotes the proportional gain. This is a fixed control strategy. The velocity of the OVF always points to the reference point, and the amplitude is determined only by the user's position.

%\vspace{-0.1cm}
\subsection{Locomotion Calculation}
%\vspace{-0.1cm}
The servo motors can measure the number of revolutions, i.e., $n_1\left(t\right)$ and $n_2\left(t\right)$, based on a rotary encoder. The locomotion distance provided by the OVF is	denoted by $\mathbf{D}^*_{\text{ovf}}(t)$, which is a linear function of $n_1\left(t\right)$ and $n_2\left(t\right)$:
%
%\begin{equation}
%\begin{split}
%\mathbf{D}^*_{\text{ovf}}(t)=\frac{d_\text{r}\pi}{\lambda}\left(
%\begin{array}{c}
% {\cos^2\theta_{\text{r1}} n_1\left(t\right)}+
%{\cos^2\theta_{\text{r2}} n_2\left(t\right)} \\
%0 \\
%{\cos\theta_{\text{r1}}\sin\theta_{\text{r1}} n_1\left(t\right)}+
%{\cos\theta_{\text{r2}}\sin\theta_{\text{r2}} n_2\left(t\right)} \\ \end{array}
%% \cos^2\theta_{\text{r1}}\frac{n_1\left(t\right)d_\text{r}\pi}{\lambda}+
%%\cos^2\theta_{\text{r2}}\frac{n_2\left(t\right)d_\text{r}\pi}{\lambda},
%%0, 
%%\cos\theta_{\text{r1}}\sin\theta_{\text{r1}}\frac{n_1\left(t\right)d_\text{r}\pi}{\lambda}+
%%\cos\theta_{\text{r2}}\sin\theta_{\text{r2}}\frac{n_2\left(t\right)d_\text{r}\pi}{\lambda}
%\right)\\
%\end{split}
%\label{eqn:Dovf}
%\end{equation}
%\vspace{-0.2cm}
\begin{equation}
%\vspace{-0.2cm}
\begin{split}
\mathbf{D}^*_{\text{ovf}}(t)\!\!=\!\!\!\left[\!\!\!\!
\Large\begin{array}{ccc}
\frac{d_\text{r}\pi\cos^2(\theta_{\text{r1}}) }{\lambda}&\!\!\!\!\!\!0\!\!\!\!\!\!&\frac{d_\text{r}\pi\cos^2(\theta_{\text{r2}}) }{\lambda}\\
0&\!\!\!\!\!\!0\!\!\!\!\!\!&0 \\
\frac{d_\text{r}\pi\cos(\theta_{\text{r1}})\sin(\theta_{\text{r1}}) }{\lambda}&\!\!\!\!\!\!0\!\!\!\!\!\!&\frac{d_\text{r}\pi\cos(\theta_{\text{r2}})\sin(\theta_{\text{r2}}) }{\lambda}  \\ \end{array}\!\!\!\!
\right]\!\!\!\!\!\left[\!\!\!\!
\begin{array}{c}
n_1\left(t\right) \\
0\\
n_2\left(t\right)\\ \end{array}\!\!\!\!
\right]\!\!\!.\!\!\!\\
\end{split}
\label{eqn:dovf}
\end{equation}
%\begin{equation}
%\begin{split}
%\mathbf{D}^*_{\text{ovf}}(t)=\left(
%\begin{array}{c}
%\frac{\cos^2\theta_{\text{r1}}d_\text{r}\pi n_1\left(t\right)}{\lambda}+
%\frac{\cos^2\theta_{\text{r2}}d_\text{r}\pi n_2\left(t\right)}{\lambda} \\
%0 \\
%\frac{\cos\theta_{\text{r1}}\sin\theta_{\text{r1}}d_\text{r}\pi n_1\left(t\right)}{\lambda}+
%\frac{\cos\theta_{\text{r2}}\sin\theta_{\text{r2}}d_\text{r}\pi n_2\left(t\right)}{\lambda} \\ \end{array}
%\right)\\
%\end{split}
%\label{eqn:Dovf}
%\end{equation}
%
%
%\begin{equation}
%\begin{split}
%\mathbf{D}^*_{\text{ovf}}(t)=\left(
%\frac{\cos^2\theta_{\text{r1}}d_\text{r}\pi n_1\left(t\right)}{\lambda}+
%\frac{\cos^2\theta_{\text{r2}}d_\text{r}\pi n_2\left(t\right)}{\lambda},
%0, 
%\frac{\cos\theta_{\text{r1}}\sin\theta_{\text{r1}}d_\text{r}\pi n_1\left(t\right)}{\lambda}+
%\frac{\cos\theta_{\text{r2}}\sin\theta_{\text{r2}}d_\text{r}\pi n_2\left(t\right)}{\lambda}
%\right)\\
%\end{split}
%\label{eqn:Dovf}
%\end{equation}
%
%\begin{equation}
%\begin{split}
%\mathbf{D}^*_{\text{ovf}}(t)=\left(
% \cos^2\theta_{\text{r1}}\frac{n_1\left(t\right)d_\text{r}\pi}{\lambda}+
%\cos^2\theta_{\text{r2}}\frac{n_2\left(t\right)d_\text{r}\pi}{\lambda},
%0, 
%\cos\theta_{\text{r1}}\sin\theta_{\text{r1}}\frac{n_1\left(t\right)d_\text{r}\pi}{\lambda}+
%\cos\theta_{\text{r2}}\sin\theta_{\text{r2}}\frac{n_2\left(t\right)d_\text{r}\pi}{\lambda}
%\right)\\
%\end{split}
%\label{eqn:Dovf}
%\end{equation}
Here $\lambda$ denotes the reduction ratio of the servo motors.
% Record the left matrix as $\mathbf{C}_T\in R^{3\times3}$, which denotes the coordinate transformation matrix of the current system.

Record the cumulative locomotion distance of the user in the physical environment and in VE as $\mathbf{D}_{\text{pe}}(t)$ and $\mathbf{D}_{\text{ve}}(t)$. Since the user walks on the OVF, part of $\mathbf{D}_{\text{pe}}(t)$ is the user’s locomotion distance that is offset by the OVF, which is denoted by $\mathbf{D}_{\text{ovf}}(t)$, and another part is the remaining distance, which is denoted by $\mathbf{D}_{\text{rm}}(t)$, Therefore,
%\vspace{-0.2cm}
\begin{equation}
%\vspace{-0.2cm}
\begin{split}
\mathbf{D}_{\text{pe}}(t)=-\mathbf{D}_{\text{ovf}}(t)+\mathbf{D}_{\text{rm}}(t).\\
\end{split}
\label{eqn:dpe}
\end{equation}

$\mathbf{D}_{\text{ovf}}(t)$ is mainly affected by the slippage during transmission. Introducing a matrix, denoted by $\mathbf{S}_L(t)\in R^{3\times3}$, to represent the impact of the sliding process, i.e.,
%\vspace{-0.2cm}
\begin{equation}
%\vspace{-0.2cm}
\begin{split}
\mathbf{D}_{\text{ovf}}(t)=\int_{0}^{t}\mathbf{S}_L(x)\mathbf{v}^{*}_{\text{ovf}}(x)dx.\\
\end{split}
\label{eqn:d*ovf}
\end{equation}

$\mathbf{v}^{*}_{\text{ovf}}(t)$ denotes the velocity of OVF and satisfies $\int_{0}^{t}\mathbf{v}^{*}_{\text{ovf}}(x)dx=\mathbf{D}^*_{\text{ovf}}(t)$. $\mathbf{S}_L(t)$ is a time-varying variable that is related to various factors, such as speed, direction and the material contacted to the platform. Since the maximum acceleration of human in daily life is about 1.44$m/s^2$ \cite{phdthesis}, even for the world 100$m$ dash champion like Usain Bolt\footnote{https://www.wired.com/2012/08/maximum-acceleration-in-the-100-m-dash/} is about 3.09$m/s^2$, an assumption in this paper is that the user has no slippage when walking on the OVF, and the following discussions based on this assumption.
%
% an reasonable assumption is the users has no slippage when walking on the OVF.
%
%This paper primarily considers an ideal circumstance, i.e., there is no slippage. 

In this case, $\mathbf{S}_L(x)$ could be simplified as a unit matrix,i.e., $\mathbf{S}_L(x)\approx \mathbf{I}_{3\times3}$, and
% $\mathbf{D}_{\text{ovf}}(t)\approx\mathbf{D}^*_{\text{ovf}}(t)$.
%\vspace{-0.2cm}
\begin{equation}
%\vspace{-0.2cm}
\begin{split}
\mathbf{D}_{\text{ovf}}(t)\approx\int_{0}^{t}\mathbf{I}_{3\times3}\mathbf{v}^{*}_{\text{ovf}}(x)dx=\mathbf{D}^*_{\text{ovf}}(t).\\
\end{split}
\label{eqn:d*ovfsimp}
\end{equation}
$\mathbf{D}_{\text{rm}}(t)$ can be measured by tracking the user's position, i.e.,
%\vspace{-0.2cm}
\begin{equation}
%\vspace{-0.2cm}
\begin{split}
\mathbf{D}_{\text{rm}}(t)=\mathbf{p}_{\text{ug}}(t)-\mathbf{p}_{\text{ref}} .
\end{split}
\label{eqn:drm}
\end{equation}

According to the control strategy, $\mathbf{D}_{\text{rm}}(t)$ can be kept within a certain range, but it does not need to be zero when stable.

Equation \ref{eqn:dpe} gives the locomotion distance in the physical environment, and the locomotion distance in VE can be obtained by the mapping process, i.e.,
%\vspace{-0.2cm}
\begin{equation}
%\vspace{-0.2cm}
\begin{split}
\mathbf{D}_{\text{ve}}(t)=\mathbf{\psi}\left(\mathbf{D}_{\text{pe}}(t),\Theta(t)\right),
\end{split}
\label{eqn:dve}
\end{equation}
Here $\mathbf{\psi}\left(x\right)$ denotes a preset function and $\Theta(t)$ denotes the external parameters and variables. $\mathbf{\psi}\left(x\right)=x$ is the simplest function, which means 1:1 mapping from $\mathbf{D}_{\text{pe}}(t)$ to $\mathbf{D}_{\text{ve}}(t)$. In addition, the $\mathbf{\psi}\left(x\right)$ could have some other formations, such as adding proportional gain, non-linear function, or even introducing the redirected-walking algorithm to get a better experience.

When the system is working, the control rate is 20$Hz$. The Algorithm\ref{algorithm:Workflow} shows the framework of the HCMK1 system. The future work to improve the performance could based on this framework.
\begin{algorithm}[htbp]
	\caption{HCMK1 Program Algorithm} %算法的名字
	\hspace*{0.02in}{\bf Input:} %算法的输入， \hspace*{0.02in}用来控制位置，同时利用 \\ 进行换行
	functions: $\mathbf{S}_L(x)$, $\mathbf{\psi}\left(x\right)$; parameters: $\theta_{\text{r1}}$, $\theta_{\text{r2}}$, $d_\text{r}$, $\lambda$, $\mathbf{p}_{\text{utrl}}$, $\mathbf{p}_{\text{ref}}$, $K_p$, $D_{th}$, $T$\\
	\hspace*{0.02in} {\bf Output:} %算法的结果输出
	locomotion distance in VE: $\mathbf{D}_{\text{ve}}(t)$
	\begin{algorithmic}[1]
		\State Initial servo motors and tracker. % \State 后写一般语句
		\State $t=0$
		\For{each cycle period with an interval of $T$,} % For 语句，需要和EndFor对应
		
		\State Receive $\mathbf{p}_{\text{tg}}(t)$ and $\mathbf{q}_{\text{tg}}(t)$ from tracker
		
		\State Calculate $\mathbf{p}_{\text{ug}}(t)$ based on Equation 3 
		\State Calculate $\mathbf{v}^*$ based on Equation 4-6
		\State Calculate $\mathbf{w}_{\text{r1}}$ and $\mathbf{w}_{\text{r2}}$ based on Equation 2
		\State Send the target rotation speed $\lambda\mathbf{w}_{\text{r1}}$ to servo motor1 and $\lambda\mathbf{w}_{\text{r2}}$ to servo motor2
		
		\State Receive $n_1\left(t\right)$ and $n_2\left(t\right)$ from servo motors
		\State Calculate $\mathbf{D}_{\text{pe}}(t)$ based on Equation 8-11
		
		\State Map $\mathbf{D}_{\text{pe}}(t)$ to $\mathbf{D}_{\text{ve}}(t)$ based on Equation 12
		\State Output $\mathbf{D}_{\text{ve}}(t)$
		
		\State $t=t+T$
		\If{Stop} % If 语句，需要和EndIf对应
		\State Quit
		\EndIf
		\EndFor
	\end{algorithmic}
	\label{algorithm:Workflow}
\end{algorithm}

%
%
%When the user walking on the OVF, the locomotion mapping from the physical world to the VE should be 1:1 with low-latency. The latency in HCMK1 includes 
%\begin{figure}[htbp]
%	\centering % avoid the use of \begin{center}...\end{center} and use \centering instead (more compact)
%	\includegraphics[width=0.91\linewidth]{pic/latency.jpg}
%	\caption{The inertial measurement unit (IMU) based positioning system. As shown in the left part image, the user stands on the omnidirectional velocity field and is tied by the belt.  There are four rotation points. Each rotation point has specific rotation axes.  }
%	\label{fig:latency}
%\end{figure}
%\begin{itemize}
%	\item $\vartriangle t_{\mathbf{v}^*}$: After setting the target velocity $\mathbf{v}^*$, the latency for the OVF to reach this velocity.
%	\item $\vartriangle t_{m}$: The delay for the servo motors to send the rotation distance to the computer and complete the coordinate transformation process.
%	\item $\vartriangle t_{tr}$: The delay for the tracker to send the position to the computer and complete the coordinate transformation.
%\end{itemize}

%
%设定速度后到最终达到该速度的反应时间，包含信号传输
%伺服电机转动距离的反馈到PC的时间
%定位器反馈到PC的时间
%
%突然停止，系统最终将人体送回到中心并停止的时间。

%
%尽管调速有延时，但该延时并不影响VR移动距离的计算
%
%移动时的距离有差距问题不大，突然停止时也能停，延时为两者最大
%
%考虑的一点问题是产生了滑动，有待进一步修正

%

%\section{naive Controller and the position calculate}

\begin{figure}[t]
%	%\vspace{-0.1cm}
	\centering % avoid the use of \begin{center}...\end{center} and use \centering instead (more compact)
	\includegraphics[width=\linewidth]{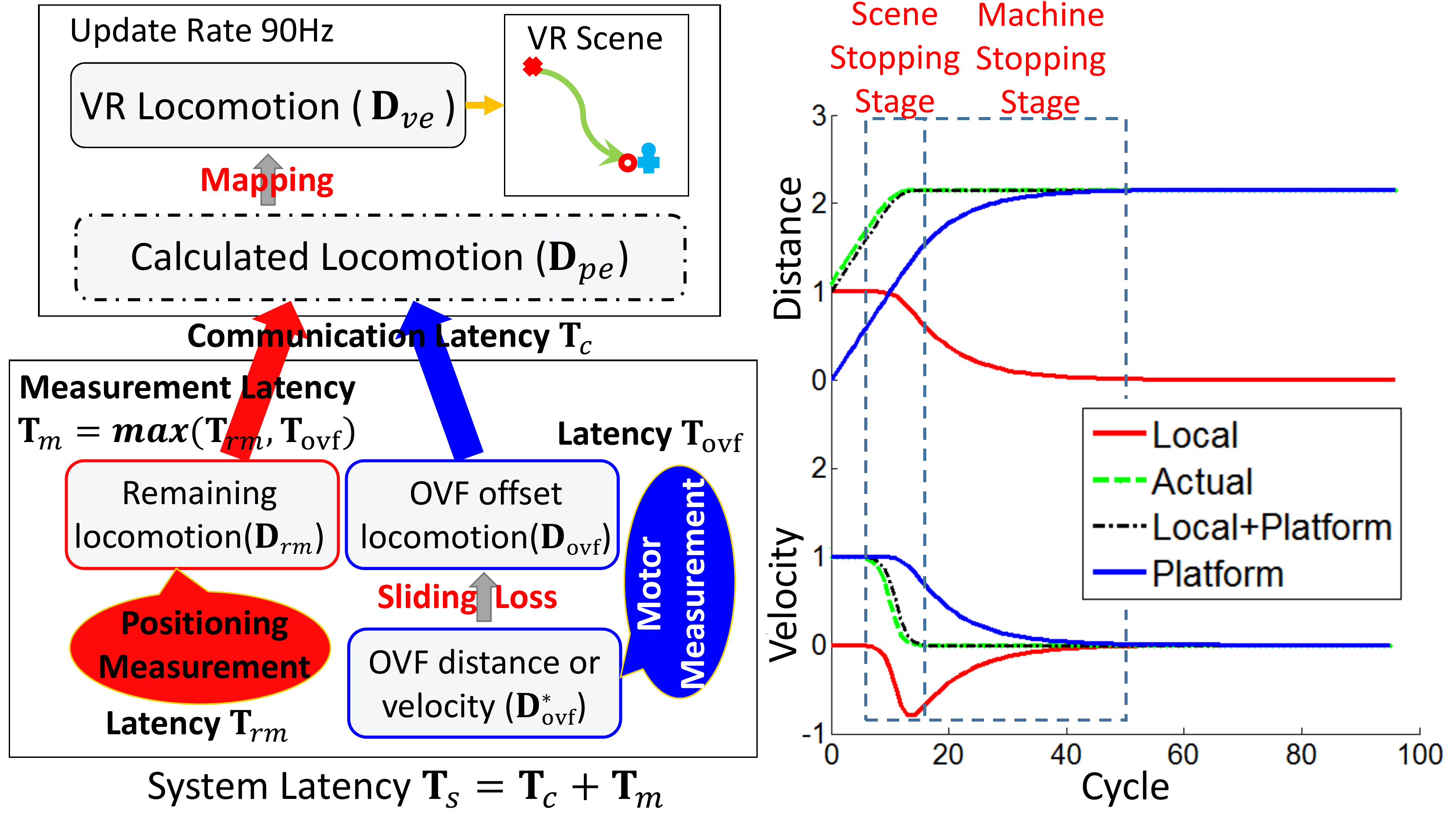}
	%\vspace{-0.6cm}
	\caption{Latency analysis of HCMK1. The left part demonstrates the latency in different measurements during the calculation process of the VR locomotion, includes the positioning latency and motor latency. The right part shows the schematic diagram when the user abruptly stops. The calculated locomotion, i.e., the black dash line, only takes a minor lag to the user's actual intended locomotion,i.e., the green dash line. 
		%		The scene can follow the user's intention to stop in the scene stopping stage. Although it takes a while for the platform to finally stop during the machine stop stage, the work delay caused by this will not affect the scene delay.
	}
	\label{fig:latency}
	%\vspace{-0.7cm}
\end{figure}

%\vspace{-0.25cm}
\subsection{Analysis of Latency}
%\vspace{-0.1cm}
Like the WiP method, HCMK1 also has a starting and a stopping latency \cite{templeman1999virtual,feasel2008llcm}, both of which are only 23ms. A notable difference is the working delay caused by the controller in HCMK1, and this delay is defined as the time needed for the device to stop working after the user stops.
Figure \ref{fig:latency} shows the sources of the latency of HCMK1. As indicated by the left part, when calculating the locomotion distance in physical environment, i.e., $\mathbf{D}_{\text{pe}}(t)$, the bottom layer needs to determine the remaining locomotion of the user by making the positioning measurement, i.e., $\mathbf{D}_{\text{rm}}(t)$, while deciding the offset distance of the OVF by making the motor measurement, i.e., $\mathbf{D}_{\text{ovf}}(t)$. The latency in the positioning measurement is about 22$ms$ \cite{niehorster2017accuracy,jones2019latency}. The motor measurement involves 2 servo motors, with RS485 serial communication bus adopted in accordance with the Modbus protocol. The latency in this process is about 8$ms$. In a multi-threaded program, the above two measurements could be performed simultaneously; therefore, the latency of the total measurement in the worst case would be $T_\text{m}=\max(T_{\text{rm}},T_{\text{ovf}})=22ms$. Then, upload the measured data to a PC through its serial port for further calculations; the communication latency in this process is $T_{\text{c}}=1ms$. In total, the maximum system latency caused by measurements and communication transmissions is $T_{\text{s}}=23ms$.

The right part of Figure \ref{fig:latency} shows the schematic diagram of each kind of data when the simple proportional controller in Subsection \ref{Basic Controller} is applied. The green dash line denotes the user’s actual intended locomotion, while the red solid line denotes the user’s local locomotion relative to the platform, i.e., $\mathbf{D}_{\text{rm}}(t)$. 
In addition, the blue solid line denotes the locomotion of the OVF on the platform, i.e., $\mathbf{D}_{\text{ovf}}(t)$.
the black dash line is the sum of the red solid line and the blue solid line, denoting the calculated locomotion in physical environment, i.e., $\mathbf{D}_{\text{pe}}(t)$.

As shown by the green dash line, at first, the user walks on the platform at a constant velocity of 1 and starts to stop at the 5th cycle. After a deceleration process, the user stops at approximately the 10th cycle, and then the actual intended velocity remains 0. The actual intended distance first increases in a linear manner, then gradually slows down, and finally remains unchanged.

Generally, this process can be divided into 4 stages, as observed in the subsequent experiment data: 
\begin{enumerate}
	\item Since the platform offsets the user’s velocity, the user’s local velocity is 0, and the local distance remains 1. The platform distance increases linearly.
	\item At the 5th cycle, the user’s actual intended velocity is decreased. Since the platform velocity is still kept at a high value, the user’s local velocity will increase but the local distance will decrease, resulting in a diminished platform velocity; accordingly, the increase of platform distance is slowed down.
	\item At the 15th cycle, the user’s local velocity equals to the platform’s velocity, the user’s actual intended velocity is decreased to 0, and the actual intended distance remains unchanged. Since the amplitudes of both the user’s local velocity and the platform’s velocity are equal but with opposite signs, the calculated velocity drops to 0, and the calculated distance no longer increases.
	\item Afterwards, the user is gradually carried back to the center by the OVF, while the user’s local distance is slowly reduced, but the platform distance is slowly increased. Finally, the platform stops working, and the user’s local velocity is reduced to 0.
\end{enumerate}

%\vspace{-0.1cm}
Assuming that the user has no slippage when walking on the OVF, a part of the user’s actual locomotion is eliminated by the OVF and the remaining part is the user’s local locomotion relative to the platform, which can be measured with millimeter accuracy by the positioning system. Therefore, the calculated locomotion is accurate, meaning that the black and green dash lines are equal in value. However, due to the system latency, the calculated locomotion always lags behind the user’s actual intended locomotion. In the worst case, the maximum latency would reach $T_{\text{s}}=23ms$.When applying the calculated locomotion to the VR scene, there is about 23$ms$ latency for which the VR scene lags behind the user’s actual intended locomotion.

Different from the 100ms VR scene latency, which is perceivable in the WiP strategy, the 23$ms$ VR scene latency of HCMK1 is so much less perceptible that it can be approximately regarded as synchronized. However, it is worth noting that the working delay caused by the controller, as shown in Figure \ref{fig:latency} for the machine stopping stage, is the time the platform continues to work after the user stops their steps. And the working delay is also caused by the inability of the simple proportional controller to accurately track the user’s actual intended velocity. Although this delay does not affect the correct and real-time mapping of the user’s actual intended locomotion to the VR scene, it will result in some discomforts, such as body shaking of users, thus leading to insecurity. In a word, how to diminish the working delay is a challenge to the control strategy.

%
%\begin{figure}[hb]
%	\centering % avoid the use of \begin{center}...\end{center} and use \centering instead (more compact)
%	\includegraphics[width=\linewidth]{pic/timing_short_n.pdf}
%	%\vspace{-0.6cm}
%	\caption{\textcolor{revised}{Timing diagram}}
%	\label{fig:timing_diagram}
%	\vspace{-0.35cm}
%\end{figure}

%
%
%When the user walking on the OVF, the locomotion mapping from the physical world to the VE should be 1:1 with low-latency. The latency in HCMK1 includes 

%\begin{itemize}
%	\item $\vartriangle t_{\mathbf{v}^*}$: After setting the target velocity $\mathbf{v}^*$, the latency for the OVF to reach this velocity.
%	\item $\vartriangle t_{m}$: The delay for the servo motors to send the rotation distance to the computer and complete the coordinate transformation process.
%	\item $\vartriangle t_{tr}$: The delay for the tracker to send the position to the computer and complete the coordinate transformation.
%\end{itemize}

\begin{figure}[hbp]
	%	%\vspace{-0.2cm}
	\centering % avoid the use of \begin{center}...\end{center} and use \centering instead (more compact)
	\includegraphics[width=0.95\linewidth]{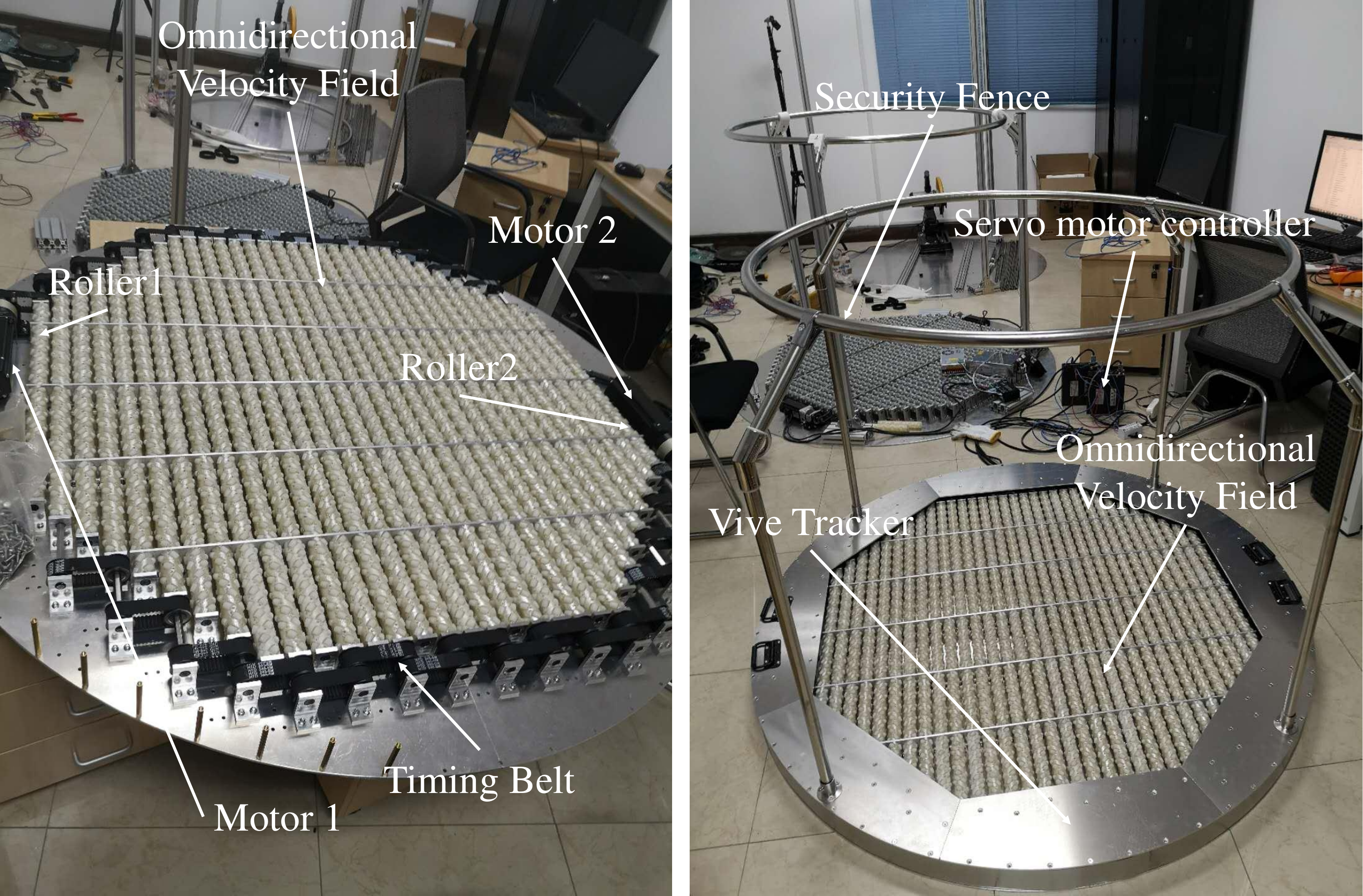}
	%\vspace{-0.3cm}
	\caption{The photographs of the HCMK1 system construction process. The OVF is consisted of 32 rollers which are driven by the two identical $0.6kW$ servo motors. The timing belts are set on the periphery to facilitate maintenance. The right part shows the overall view of the HCMK1 system. The height is only $8cm$ and the diameter is $1.5m$. The total weight, including the security fence, is $110kg$. }
	\label{fig:Systemoverview}
	%\vspace{-0.3cm}
\end{figure}
%\vspace{-0.2cm}
\section{High-Level Comparisons With Other Systems}
\begin{figure}[tbp]
	\centering % avoid the use of \begin{center}...\end{center} and use \centering instead (more compact)
	\includegraphics[width=\linewidth]{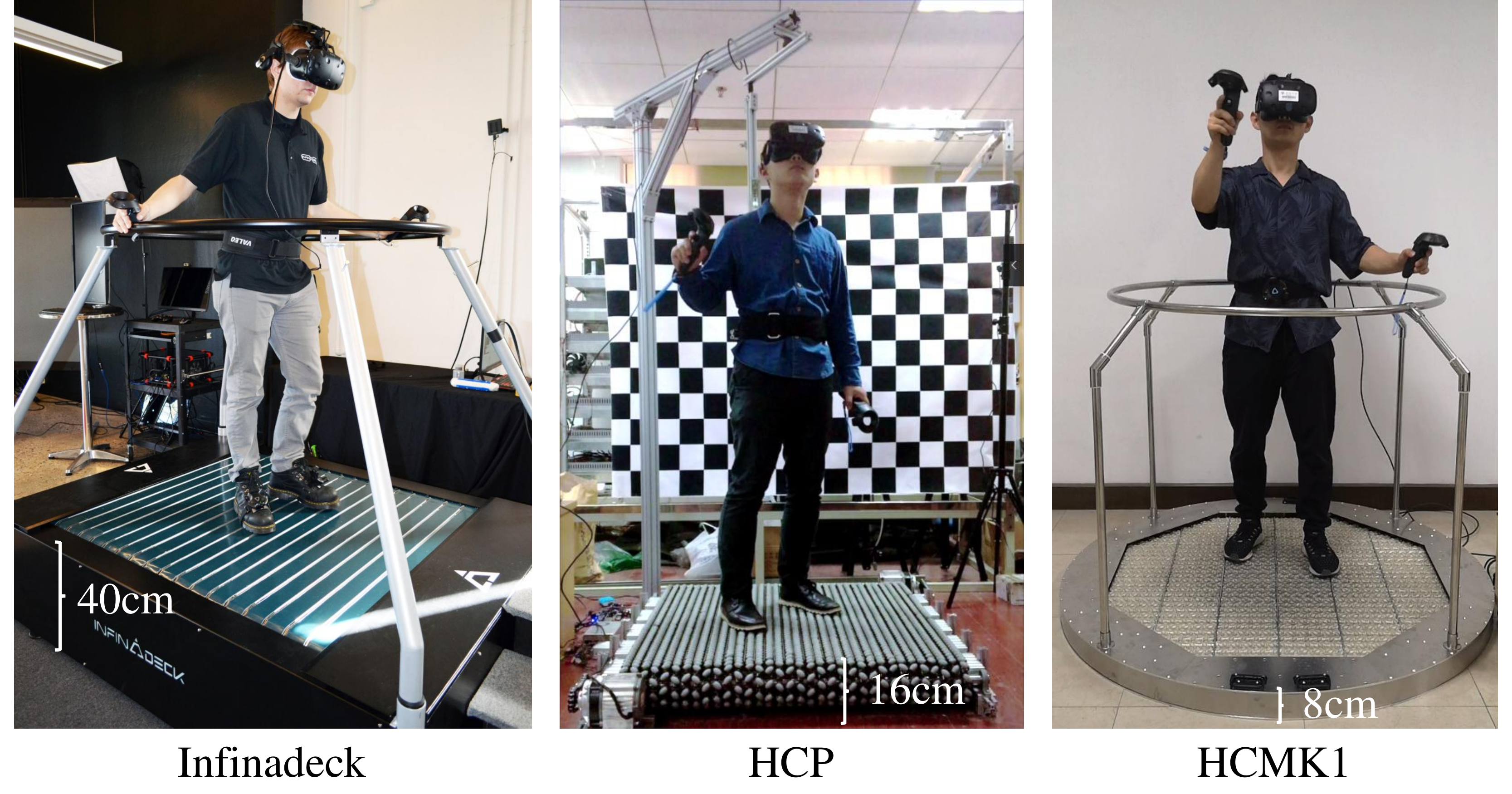}
	%\vspace{-0.7cm}
	\caption{The comparison of HCMK1 with Infinadeck and HCP system. HCMK1 is much more compact, and has only 20\% of the height of Infinadeck and 50\% of the HCP. It means the corresponding volume will reduce at least 80\% and 50\% respectively. }
	\label{fig:comparison}
	%\vspace{-0.7cm}
\end{figure}

A working HCMK1 system is shown in Figure \ref{fig:Systemoverview}. Since the OVF should be isotropic, a circular structure is more reasonable than a square structure. The mirror-symmetrical spiral rollers make it easier to design the platform close to a circle. In HCMK1, the OVF is designed as a regular octagon, and it saves the waste area on the diagonal of the square structure. Figure \ref{fig:comparison} shows the comparison of HCMK1 with the Infinadeck and the HCP system. It shows HCMK1 is a much lighter device and is more suitable for the room-scale VR.

\begin{table*}[htb]
	\caption{Comparison with different omnidirectional treadmills.}
	\scriptsize%
	\label{tab:comparison}
	\centering
	%\vspace{-0.1cm}
	\begin{tabular}{cccccccc}
		\toprule
		Device                      & \multicolumn{2}{c}{Motor} & Active Area                    &   Height                 & Weight                   & Maximum Speed & Maximum Acceleration                       \\ \midrule
		\multirow{2}{*}{CyberWalk}  & X-axis      & 40kW(4EA)        & \multirow{2}{*}{6.5*6.5m} & \multirow{2}{*}{1.45m} & \multirow{2}{*}{12000kg} & \multirow{2}{*}{2m/s} & \multirow{2}{*}{0.75m/s$^2$}              \\ 
		& Y-axis      & 37.7kW(25EA)      &                           &                        &                          &                                     \\ 
		\multirow{2}{*}{F-ODT}      & X-axis      & 8.8kW(2EA)       & \multirow{2}{*}{2.5*2.5m} & \multirow{2}{*}{0.64m} & \multirow{2}{*}{576kg}   & \multirow{2}{*}{2.5m/s} & \multirow{2}{*}{3m/s$^2$}            \\
		& Y-axis      & 3.6kW(2EA)       &                           &                        &                          &                                     \\ 
		\multirow{2}{*}{Infinadeck} & X-axis      & -           & \multirow{2}{*}{1.2*1.5m} & \multirow{2}{*}{0.4m}  & \multirow{2}{*}{225kg}   & \multirow{2}{*}{\textgreater{}2m/s}& \multirow{2}{*}{-} \\ 
		& Y-axis      & -           &                           &                        &                          &                                     \\ 
		\multirow{2}{*}{HCP}        & X-axis      & 0.5kW(1EA)       & \multirow{2}{*}{1*1.2m}  & \multirow{2}{*}{0.16m} & \multirow{2}{*}{150kg}   & \multirow{2}{*}{1.6m/s}& \multirow{2}{*}{1.3m/s$^2$}             \\ 
		& Y-axis      & 0.5kW(1EA)       &                           &                        &                          &                                     \\
		\multirow{2}{*}{\bf HCMK1}        & X-axis      & 0.6kW(1EA)       & \multirow{2}{*}{Radius 0.575m, 1.10$m^2$}  & \multirow{2}{*}{0.08m} & \multirow{2}{*}{110kg}   & \multirow{2}{*}{1.78m/s}& \multirow{2}{*}{25.00m/s$^2$}             \\ 
		& Y-axis      & 0.6kW(1EA)       &                           &                        &                          &                                     \\ \bottomrule
	\end{tabular}
	%\vspace{-0.4cm}
\end{table*}

The detailed parameters of HCMK1 is shown in Table \ref{tab:parameter}. We applied two 0.6$kW$ servo motors and the gear ratio is 3:1. The rotation part of HCMK1, i.e., the spiral rollers and synchronous wheels, is about 48$kg$ weight totally and the rotation radius is 1.71$cm$. Therefore, the moment of inertia is only 0.007$kg\cdot m^2$. The small moment of inertia ensures that even applying low power motors can obtain sufficient dynamic performance. Theoretically, when assuming the transmission efficiency is 90$\%$, the maximum acceleration can reach 25.00$m/s^2$,  and for a user with 100$kg$ weight, the acceleration can reach 4.84$m/s^2$. 
The experiment in the next section 
%The additional experiment in supplementary material
shows the maximum starting acceleration is about 16.00$m/s^2$ and the maximum braking acceleration can reach 30.00$m/s^2$. 
The transmission efficiency is about 88\% and the torque caused by rotating
friction is about 2.5NM.

Table \ref{tab:comparison} demonstrates the detail parameters of several systems that can generate parallel OVF. Cyberwalk \cite{souman2011cyberwalk,schwaiger2007cyberwalk}, F-ODT \cite{lee2016design,pyo2018development}, Infinadeck \cite{metsis2017integration} are three belt-based ODTs. Huge volume and weight lead to large inertia, which puts much more pressure on the actuator and reduces the dynamic performance. To produce enough power, Cyberwalk and F-ODT set several motors at one axis. Therefore, they need to solve synchronization errors\cite{pyo2018development} between different motors. In addition, the motors in X-axis and Y-axis are different and always have different electrical characteristics, such as the rated speed, the torque etc. All these will bring difficulties to the controller.

\begin{table}[tbp]
	\caption{The detail parameters of HCMK1 system.}
	\scriptsize%
	\label{tab:parameter}
	\centering
	%\vspace{-0.1cm}
	\begin{tabular}{lclc}
		\toprule
		\multicolumn{4}{c}{Motor Model}  \\ \midrule
		Rated Power     & 0.6 kW         & Torque              & 1.9 NM            \\ 
		Rated Speed     & 3000 rpm       & Gear Ratio            & 3:1            \\ \midrule
		%		Gear Ratio      & \color{red}3:1          & Update Rate           & 500hz          \\ \midrule
		\multicolumn{4}{c}{Device Parameter}                                     \\ \midrule
		Total Power     & 1.2kW           &Total Area      & 1.76$m^2$       \\ 
		Height         &  0.08m   &   Weight                & 110kg        \\ 
		Maximum Speed    &  1.78m/s    & Maximum Acceleration     & {25.00m/s$^2$}          \\ 
		Active Area     & 1.10$m^2$     &    User Acceleration (100kg)              & {4.84m/s$^2$}          \\ \bottomrule
	\end{tabular}
	%\vspace{-0.65cm}
\end{table}

HCP \cite{Wang2020} applied the 45-degree wheel-based scheme and reduced the volume and weight a lot based on the mirror-symmetrical chains. HCMK1 proposes a novel carrier, i.e., the mirror-symmetrical spiral rollers, to replace the chains in HCP. It significantly reduces the device weight and moment of inertia. Therefore, although the total power is only 1.2$kW$, the dynamic performance can be greatly improved. Furthermore, the mirror-symmetrical structure ensures the identical motors can be applied in different axes.

\begin{figure*}[htbp]
	\centering % avoid the use of \begin{center}...\end{center} and use \centering instead (more compact)
	\includegraphics[width=1\linewidth]{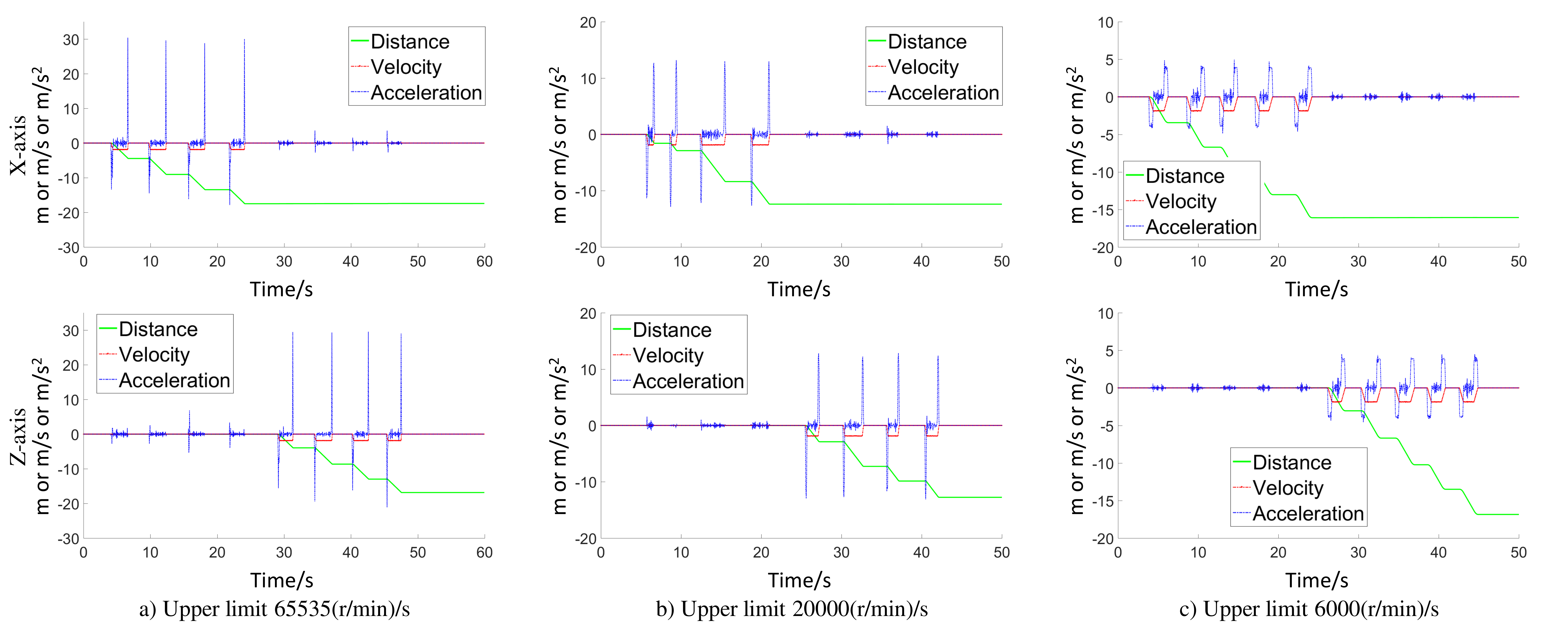}
	\vspace{-0.5cm}
	\caption{ Three acceleration experiments. Each column demonstrates the recorded data of the X and Z axes with different upper limits of the servo motors' acceleration. For each experiment, we conduct four starting-braking operations along each axis. The collected data includes the distance, velocity and acceleration of the platform. }
	\label{fig:experiment_acceleration}
	\vspace{-0.2cm}
\end{figure*}

\begin{figure*}[h]
	\centering % avoid the use of \begin{center}...\end{center} and use \centering instead (more compact)
	\includegraphics[width=1\linewidth]{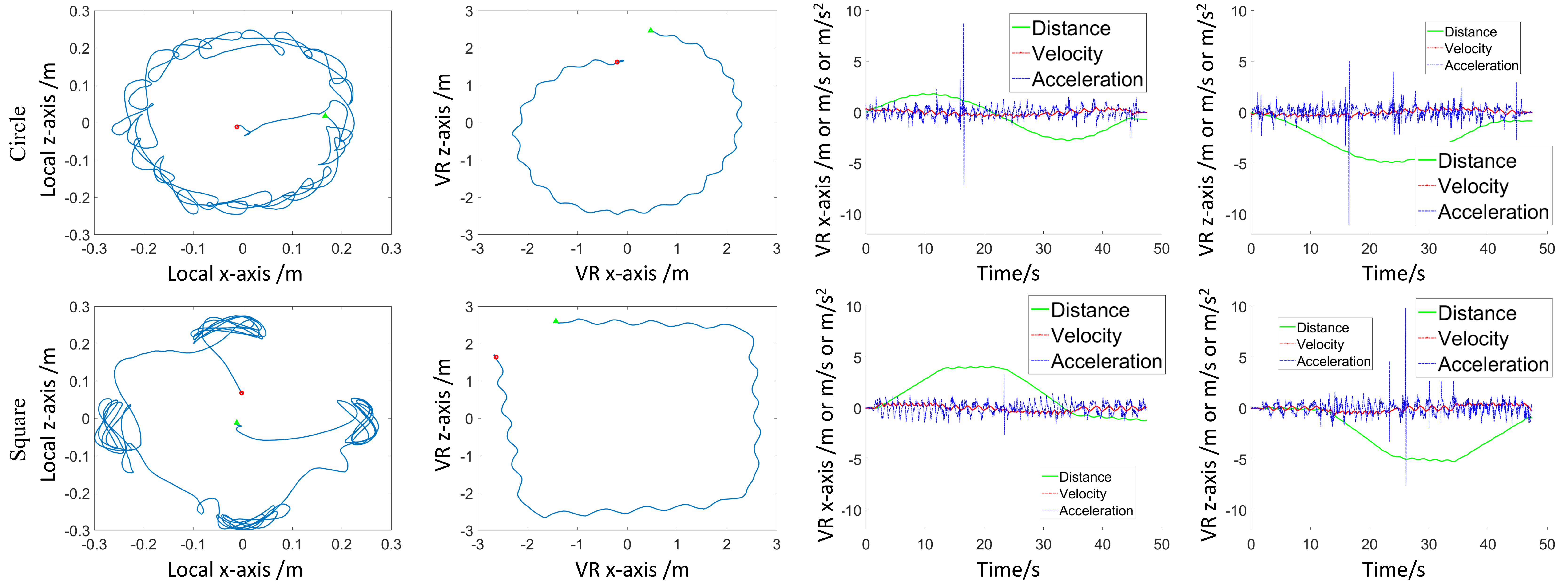}
	\vspace{-0.5cm}
	\caption{User's locomotion experiments. Each row represents an experiment in which the user walks along a certain trajectory. The first column is the user's local position relative to the platform. The second column is the user's VR position relative to the coordinate system in VE. The green point denotes the start point and the red point denotes the endpoint. The third and fourth columns represent the corresponding spatial information in the VE coordinate system along the X and Z axis respectively. }
	\label{fig:experiment_locomotion}
	\vspace{-0.3cm}
	%\vspace{-0.6cm}
\end{figure*}

\section{Experiment Result}
%\vspace{-0.1cm}
%All the experiments 
%3d position data 
%the speed Analysis
%the main factor of user experience
\subsection{Acceleration Experiment}
Sufficient dynamic performance is the basis for designing a good controller. Given HCMK1 delivers a small moment of inertia, as described in Section 4, the platform can theoretically achieve an acceleration of 25.00$m/s^2$. In this subsection, several experiments are made to test the actual acceleration performance. The experiments record the starting and braking process from 0 to the maximum speed, i.e., 1.78$m/s$, without additional load imposed, but with different upper limits of the motor acceleration applied.

Figure \ref{fig:experiment_acceleration} demonstrates the results of these experiments. The first one mainly tests the maximum acceleration of the platform. The upper limit is set to 65535$(r/min)/s$, corresponding to 39.32$m/s^2$ on the surface of the OVF, a figure that is much greater than the theoretical value of 25.00$m/s^2$. As indicated in Figure \ref{fig:experiment_acceleration} (a), the starting acceleration is about 16.00$m/s^2$ and the braking acceleration is about 30.00$m/s^2$. It may be caused by the rotating friction. From this result, it can be calculated that the transmission efficiency is about 88$\%$ and the torque caused by rotating friction is about 2.5$NM$. 

In the second experiment, as demonstrated in Figure \ref{fig:experiment_acceleration} (b), when the upper limit is set to 20000$(r/min)/s$, i.e., 12.00$m/s^2$, the starting and braking accelerations of OVF are almost the same, about 12.5$m/s^2$, indicating that the OVF is fully controllable in this range. For the design of high-level controllers, the driving force is sufficient to directly control the acceleration, while ignoring the rotating friction.

The third experiment implements the upper limit of  6000$(r/min)/s$, i.e., 3.6$m/s^2$. Figure \ref{fig:experiment_acceleration} (c) indicates that the starting and braking accelerations are both around 3.8$m/s^2$ which is more stable than in the second experiment. Actually, excessive acceleration will cause users to lose balance; therefore, limiting the acceleration to a certain range is essential for improving the UX.

Figure \ref{fig:experiment_acceleration} further demonstrates that the X and Z axes have the same dynamic performance, which results from the mirror-symmetric spiral roller structure of HCMK1.

\subsection{Locomotion Experiment}
%\vspace{-0.1cm}
This experiment is designed to verify the effectiveness of the system in terms of space expansion, participated in by a skilled male experimenter with a height of 174$cm$ and a weight of 72$kg$ participated in this test. The locomotion experiment includes two sub-experiments, in which the user will try to walk along a circular trajectory and a square trajectory on the platform, with the recorded data shown in Figure  \ref{fig:experiment_locomotion} where, the first row shows the circle trajectory’s experiment result and the second row shows the square trajectory’s experiment result. The user starts to move at the green point and stops at the red point. The fluctuations in the trajectory are caused by the swings of the human body when moving. The results in all of the experiment prove that the platform can effectively expand the limited physical locomotion to limitless VR locomotion.

\begin{figure*}[t]
	\centering % avoid the use of \begin{center}...\end{center} and use \centering instead (more compact)
	\includegraphics[width=\linewidth]{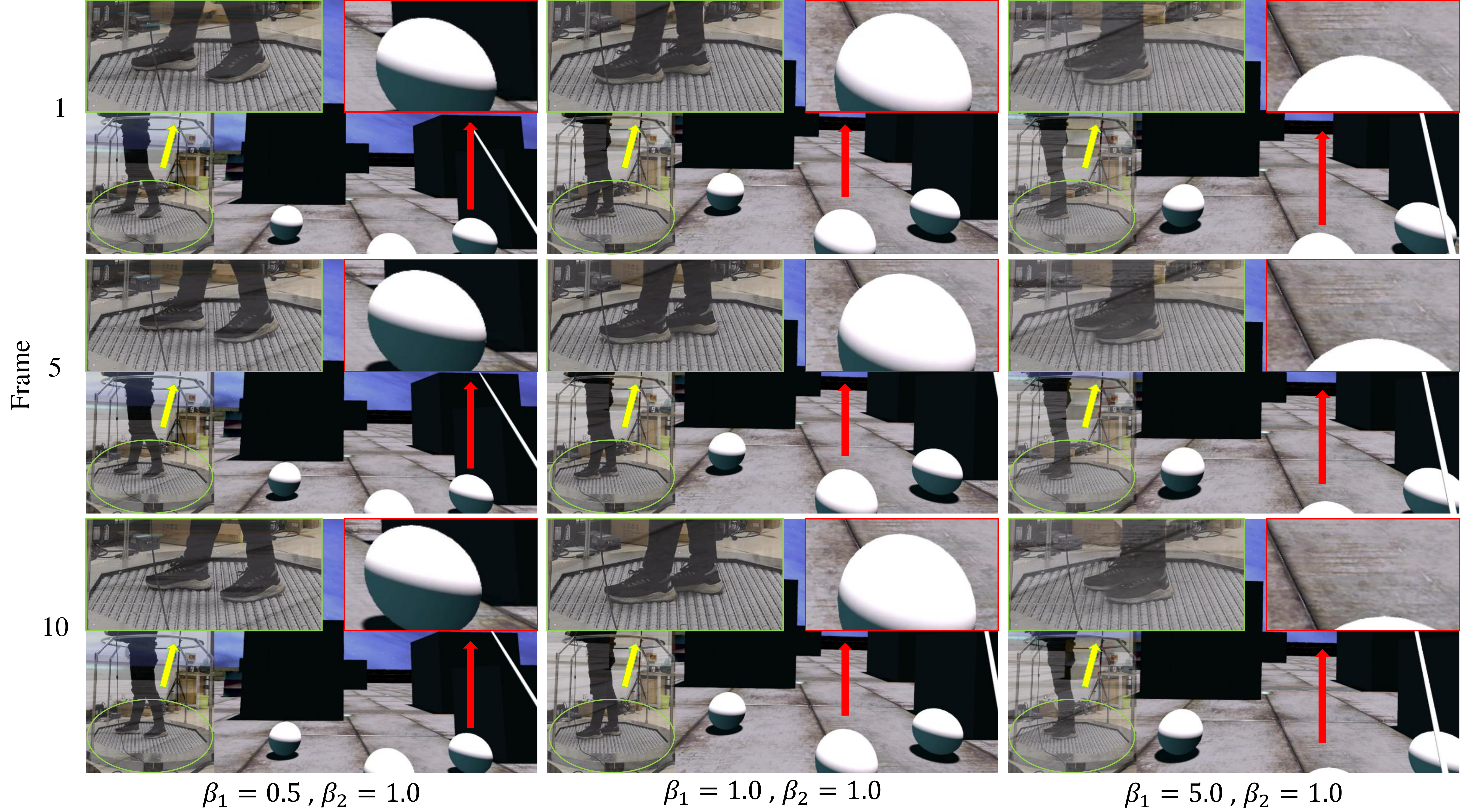}
	%\vspace{-0.8cm}
	\caption{The screenshots of the process from the user stops until the platform carries the user back to the center. Each column applies a different gains ratio. The green and red boxes show the zoomed part of the VR scene, includes the position of the user on the platform and the salient reference objects. It is obvious that, after the user stops, the VR scenes slide backward in the left column, stop immediately in the middle column and slide forward in the right column  }
	\label{fig:experiment_latency}
	%\vspace{-0.65cm}
\end{figure*}

\begin{figure}[t]
	\centering % avoid the use of \begin{center}...\end{center} and use \centering instead (more compact)
	\includegraphics[width=0.725\linewidth]{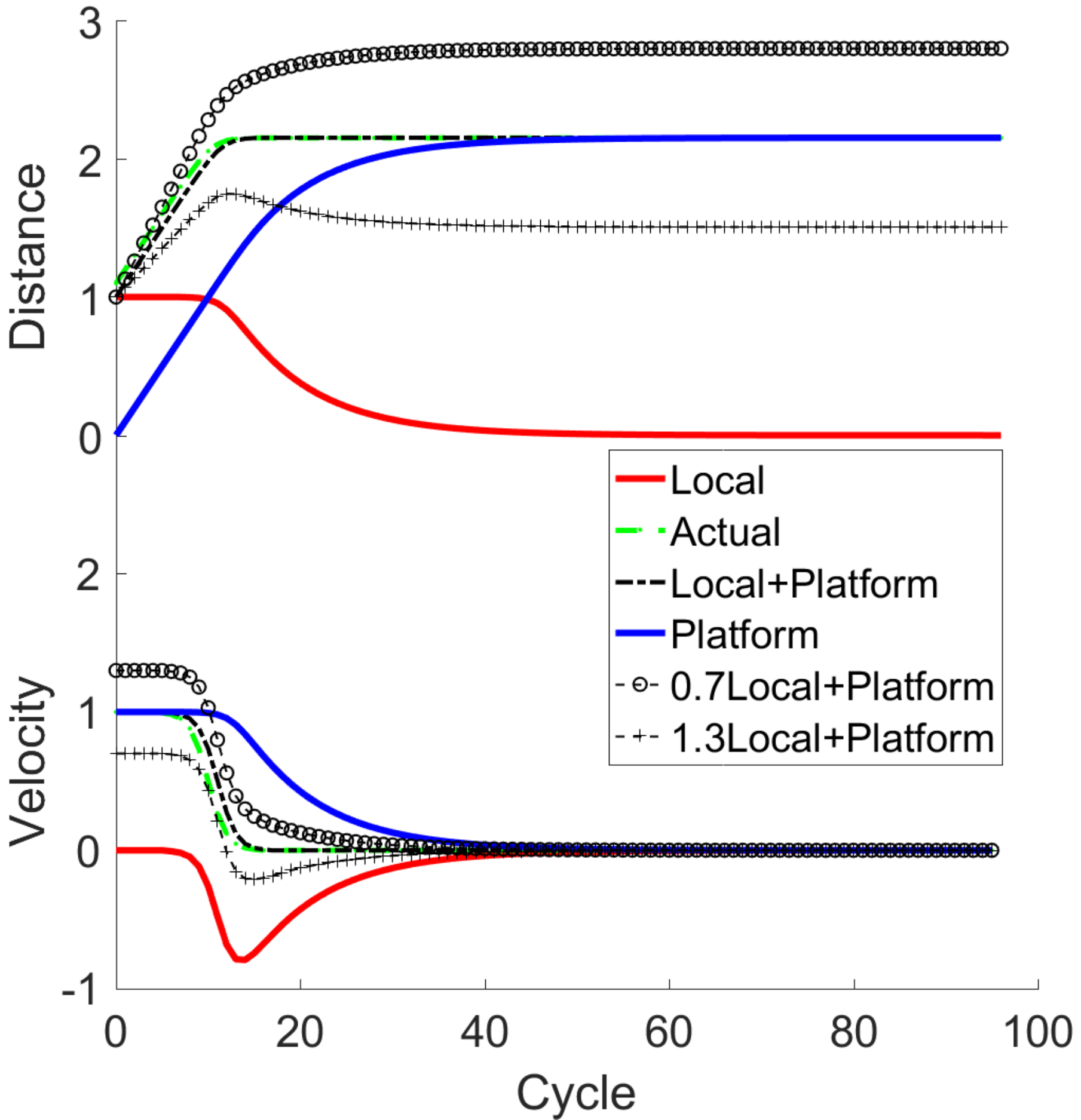}
	%\vspace{-0.3cm}
	\caption{The schematic diagram of different gains ratios. If the gain of the user's local locomotion mismatches the gain of the platform locomotion, the working delay will increase the scene latency, causing the VR scenes to slide backward or forward.}
	\label{fig:latency_gains}
	%\vspace{-0.6cm}
\end{figure}

%\vspace{-0.1cm}
\subsection{Locomotion Gains and Scene Latency \label{latency_experiment}}
%\vspace{-0.1cm}
It is found in the latency analysis that the VR scenes and the user’s actual intentions are almost synchronized. That is based on the Equation \ref{eqn:dpe}, i.e., the gains of the platform OVF locomotion and the user’s local locomotion is 1:1. The mutual elimination of these two sets of locomotion ensures that the working delay would not affect the scene latency. As described in Equation \ref{eqn:dve}, changing the mapping function can introduce different algorithms or achieve special effects. However, the ratio of gains should be 1:1 in this mapping function; otherwise, the working delay will enhance the scene latency.

In this experiment, we use
%\vspace{-0.2cm} 
\begin{equation}
%\vspace{-0.2cm}
\begin{split}
\mathbf{D}_{\text{pe}}(t)=-\beta_{1}\cdot\mathbf{D}_{\text{ovf}}(t)+\beta_{2}\cdot\mathbf{D}_{\text{rm}}(t).\\
\end{split}
\label{eqn:dpetheta}
\end{equation}
to analyze the influence of different gains on the scene latency. Based on Figure \ref{fig:latency_gains}, the following simple inferences can be acquired:
\begin{enumerate}
	\item When the user stops, if $\beta_{1} < \beta_{2}$, the locomotion of platform OVF cannot completely eliminate the user’s local locomotion; therefore, the VR scene would slide in the backward direction; 
	\item If $\beta_{1} = \beta_{2}$, the locomotion of platform OVF would eliminate the user’s local locomotion exactly; therefore, the VR scene will stop synchronously;
	\item If $\beta_{1} > \beta_{2}$, the locomotion of platform OVF would excessively eliminate the user’s local locomotion, therefore, the VR scene will continue to slide in the forward direction.
\end{enumerate}

One participant in this experiment has a height of 174$cm$ and a weight of 72$kg$. The HMD is HTC Vive. Since human bodies will inevitably shake when the platform is working, in order to relieve the impact of body shaking on VR scenes, the experimenter is required to keep stable after stopping his moves.  
%Since the 6DoF positioning of the HMD was completed by the SteamVR SDK, the default positioning data was not modified in the experiment,i.e., $\beta_{2}=1.0$ .
The ratio between $\beta_{1}$ and $\beta_{2}$ is set as 0.5:1, 1:1 and 5:1.
By selecting the data with less body shaking, the results in Figure \ref{fig:experiment_latency} are obtained. The frame rate is 29.97$Hz$. The 1st, 5th and 10th frames of the user’s stopping process are extracted. The continuous experimental video is attached in the supplementary material.

Different gain ratios are applied in different columns in Figure \ref{fig:experiment_latency}. The positions of the user on the platform and the salient reference objects in the scenes are zoomed in. It can be clearly observed that when $\beta_{1} \neq \beta_{2}$, the VR scenes’ stops always lag significantly behind the user’s stop actions. This finding is consistent with the inference above: when $\beta_{1}<\beta_{2}$, the VR scenes would slide backward; when $\beta_{1}>\beta_{2}$, the VR scenes would continue to slide forward; and when  $\beta_{1}=\beta_{2}$, the VR scenes would stop immediately.

This experiment demonstrate that when calculating a user’s locomotion or introducing mapping functions as Equation \ref{eqn:dve}, it is necessary to ensure that the gains of the platform OVF locomotion and the user’s local locomotion are equal. Otherwise, it will bring about a large scene latency between the VR scene and the user’s intention.

\begin{figure*}[t]
	\centering % avoid the use of \begin{center}...\end{center} and use \centering instead (more compact)
	\includegraphics[width=1\linewidth]{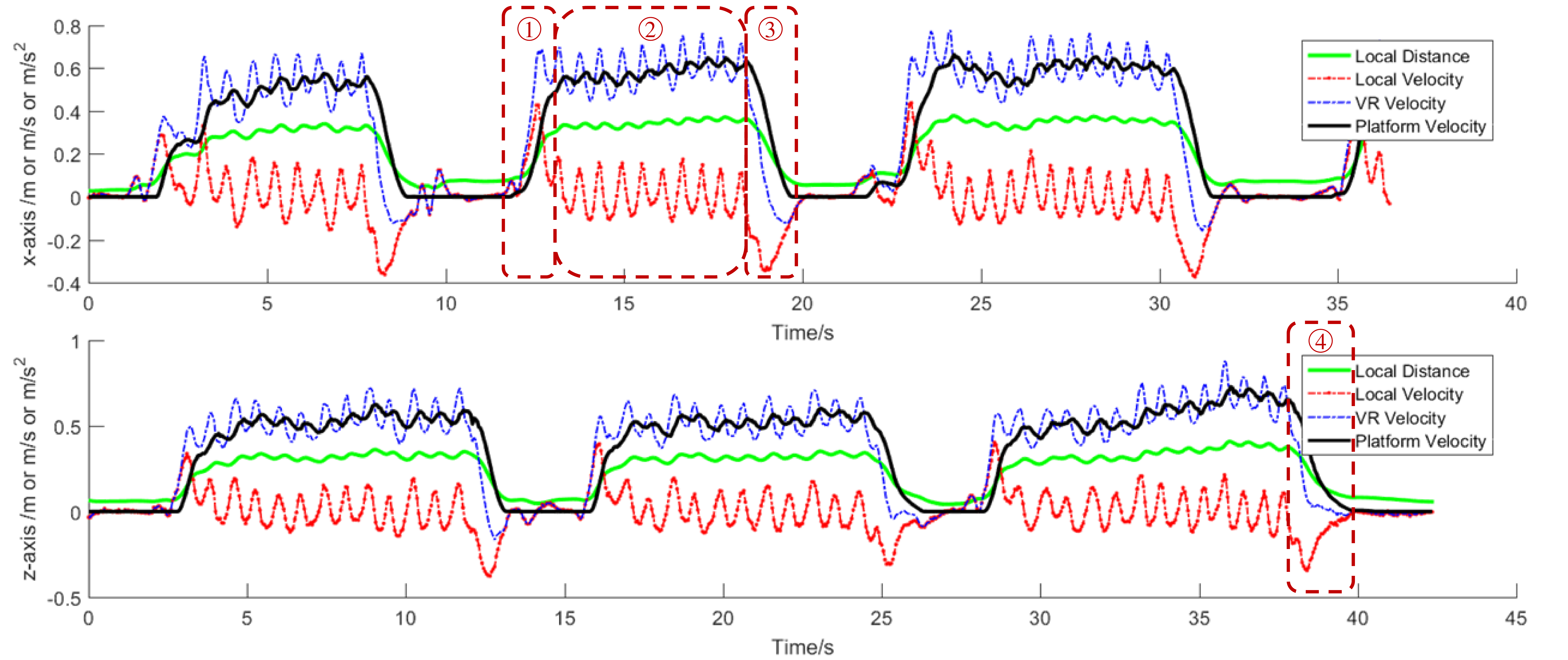}
	%\vspace{-0.7cm}
	\caption{Each row demonstrates the spatial information of the user walking along different axes on the platform. 
		%		The local distance, local velocity, and VR velocity are the spatial information about the user. The platform velocity is the spatial information about the platform. 
		Each experiment contains 3 stages, among which, the beginning and the end are the main stages that affect the UX.}
	\label{fig:experiment_performance}
	%\vspace{-0.6cm}
\end{figure*}

%\vspace{-0.15cm}
\subsection{Main Factors Affecting Users’ Experience}
%\vspace{-0.1cm}
At present, a total of 30 participants (including 5 females and 25 males) have experienced the VR interaction on this system. Age range: 20 to 45 (27.67$\pm$7.68).  Height range: 155 to 198$cm$ (172.03$\pm$8.01$cm$). Weight range: 45 to 99$kg$ (67.82$\pm$12.73$kg$). All of the participants were first-time users. After walking on the treadmill for about 1$min$ without wearing HMD, they then entered the same VR scenes to have a free interaction. Aside from reminding about the security handrail, no other guidance or instructions are given. The average experiencing time is 10 minutes. An example of the interaction process is shown in the attached application demo video.
%The VR scene contains blocks, oil drums, walls, and other obstacles. The user can hold a stick on one hand which can be used to poke nearby obstacles and generate bombs on the other hand.  The bombs can be thrown away and when it hits an object, it will explode and knock nearby objects away. 
After completing the experience, each of the participants were inquired whether they had experienced motion sickness or any other discomforts.

The inquiry finds that 29 participants acquired good experiences without motion sickness, with only one encountering slight motion sickness, who is 198$cm$ tall and weight 99$kg$. Besides, during the experiments, 4 participants attempted to release the security handrail when walking. Among them, two found it was hard to stabilize their bodies when stopping and turning around, with a feeling of falling down; another two indicated no abnormalities. None of these 4 participants experienced motion sickness.

According to the above results, the UX needs to be further improved. Although HCMK1 has many advantages in its design, like other driven-based ODTs, the UX would ultimately depend on the performance of the controller \cite{de2009control,asl2018intelligent}. This is a human–computer interaction problem. The ideal UX is that no matter when a user starts or stops moving at any speed, his/her body can always keep stationary at the center of the platform. Obviously, there are still a certain gap between the current experiences and the ideal cases. To further pinpoint the causes of the above results to boost the controller’s performance, this study collects and analyzes the actual data through additional experiments.

The experiments include two independent start-stop movements along the X-axis and Z-axis by one participant at a height of 174$cm$ and weight of 72$kg$, wearing HTC VIVE HMD and tracker. The data sampling rate is 20$Hz$. The participant was asked to accelerate from a stationary state, walk at a constant speed for a while, and then immediately stop steps and remain stationary until the platform stops running. During the whole process, the user moved without the help of handrails. Unlike the scene latency experiment in Subsection \ref{latency_experiment}, in this experiment, participant was only asked to walk naturally, without deliberately keeping the body stable. The data collected include the user’s local distance and velocity, the platform OVF velocity, and the calculated VR velocity. The calculation process is based on Equation \ref{eqn:dpe}, i.e., $\beta_{1}=\beta_{2}=1.0$.

Figure \ref{fig:experiment_performance} demonstrates part of the continuously recorded experiment data. The start-stop process corresponds to the three dark-red dash-line boxes, i.e., \textcircled{1},\textcircled{2},\textcircled{3}. Since the results about the X and Z axes are almost the same, the discussion will be mainly focused on the X-axis.

In the first stage, i.e., Box \textcircled{1}, when the user starts walking but within the threshold of the controller, the platform keeps stationary. The user’s local velocity increases until the user’s position exceeds the threshold. Then, the platform starts to carry the user to go backwards and thus slows down the user’s local velocity; however, the user’s VR velocity still rises as indicated by the actual intention, therefore leading to the peak of the red dash-line. 

When the user’s local velocity decreases to approach 0, the second stage starts and the user’s actual intended velocity is eliminated by the platform velocity. When the user steps, the local velocity presents fluctuations, but the average value is approximate to 0.

The third stage starts when the user suddenly stops. The local velocity is lowered to a negative value due to the influence of the platform velocity. Then the user enters the threshold scope and the platform stops working. 
It is worth noting that, although the peak of VR velocity, as shown by the blue dashed line in Box \textcircled{3} resembles the case of  $\beta_{1}<\beta_{2}$ in Figure \ref{fig:latency_gains}, when the OVF velocity decreases to 0, the user’s local velocity is not 0. From this finding, it can be inferred that the peak is not caused by the unmatching gains in the calculation process, but because of the shaking of body. In contrast, in Box \textcircled{4}, the user keeps the body shaking at a low level during the stop stage. When the platform OVF velocity falls to 0, the user’s local velocity becomes substantially 0, while the calculated VR velocity drops to 0 earlier than the platform.

In both X-axis and Z-axis, it can be observed that the shaking of body during the stop stage is a common problem. Due to the working delay, the platform OVF velocity fails to track the user’s intended velocity quickly. After the user stops, the platform has to take some time to send the user back to the center. Meanwhile, the center of gravity of the user is far above the ground. As a result, after the platform sends the user back to the center and stops working, the body still moves at a certain speed due to inertia, which leads to shakes of body. Besides, the higher the user’s center of gravity, the more intense the shaking phenomenon. Once the shakes reach a certain extent, the sense of motion sickness will appear. The fall-down feeling appearing when stopping and turning without holding the handrail can also be explained with the same reason. In addition, physiological studies reveal that humans are much more sensitive to lateral acceleration \cite{de2009control}; therefore, body shaking caused by the working delay will become even more serious when turning around.

Although previous researches \cite{asl2018intelligent} have found that the main factor affecting the UX is users’ position errors in the abrupt stop stage. This study maintains that such errors are just one of the main reasons affecting the UX, and other reasons also include the system’s working delay and the height of the center of gravity. The distance for OVF to send users back to the center can be shortened by cutting users’ local distance in Stage \textcircled{2}. Therefore, minimizing the initial position errors of users during the abrupt stop stage can lower the working delay for a while. Another method to reduce the working delay is to predict users’ stopping action in advance and set the platform OVF velocity to 0, so that the platform OVF can quickly follow the users to stop. In addition, in order to obtain a good UX, body shaking caused by inertia is unacceptable. Therefore, when designing the controller, it is necessary to consider inhibiting the influence of the height of the center of gravity.
\section{Limitations}
%\vspace{-0.1cm}

Although HCMK1 is superior to other similar systems, it still has some limitations with respect to the mechanism, algorithms and the metrics of UX. These limitations may limits the possible application scenarios of HCMK1.

We noticed that the working noise of HCMK1 was relatively large, which might have a negative effect on the UX.  The noise was mainly caused by the power supply fan and mechanical transmission.  Moreover, the security handrail might negatively affect the user to use the controller for VR interaction. 
%
%In terms of the hardware system, the electromagnetic interference(EMI) generated by the 220$V$ AC servo motor will cause the HMD unable to positioning. In contrast, the wirelessly connected HTC VIVE TRACKER doesn’t have this problem. In order to solve this problem, we have tried to add isolation power supply, power supply filter and magnetic ring. However, none of them can completely solve this problem. After replacing the 220$V$ AC servo motor of another manufacturer, the EMI is alleviated, but it still exists. Finally, we decide to use a 48$V$ low-voltage DC servo motor. At present, the problem that the HMD can’t be positioned due to interference is basically solved. However, the reason why the AC servo motor produces the EMI is still not found, which needs further study.
%
%As a general-purpose input platform, but the current software system has insufficient versatility, which only supports the development of Unity and lacks support for other platforms. There is also only one scene available for experience.

The control algorithm is the key problem that needs to be studied and solved urgently in the current system. This paper simply uses the proportional control for verification, whose final performance is not satisfactory. Because the large working delay might have a negative impact on the UX, the proportional control algorithm that we used in HCMK1 should be replaced by better control algorithms for optimal performances.

In addition, current evaluation of UX only stays at a subjective level, that is, the judgment is made by inquiries. The lack of a quantitative metric leads to it is difficult to distinguish different UX brought by different algorithms.
% Although the main factors affecting UX are obtained by analyzing the measured data, there is a lack of a single quantitative metric that can effectively evaluate the UX, therefore, it is difficult to quantitative distinguish different UXs brought by different algorithms.
%\vspace{-0.2cm}
\section{Conclusions and Future Work}
%\vspace{-0.1cm}
%This paper proposes and realizes a kind of brand-new ODT, namely HCMK1 for the current VR industry bottlenecks of natural locomotion interface problems. The scheme by 45-degree wheels velocity synthetic method, can not only provide a natural walking experience but also has a great advantage in the aspects of weight, volume, dynamic performance, and scene latency, making it possible to achieve real immersion in Room-Scale VR. Through the experiments and analysis of latency, this paper systematically verifies that when the gains of OVF’s locomotion and the user’s local locomotion are matching, HCMK1 has a minor scene delay, only 23$ms$. By analyzing the subjective results and objective data of the UX, this paper summarized the influence of the controller on the final UX, among which the main factors are the working delay and the height of the center of gravity. This is helpful for designing better controllers to improve UX in the future.

This paper has proposed and developed a novel ODT system, namely HCMK1, suitable for household VR applications. The design scheme not only provides a natural walking experience but also has great advantages with respect to the weight, volume, dynamic performance, and scene latency, making it possible to achieve real immersion in Room-Scale VR.  Through the experiments and analysis, this paper has systematically validated that when the gains of OVF’s locomotion and the user’s local locomotion match, HCMK1 has only a minor scene delay of 23$ms$. By analyzing the results of the UX, this paper has identified several main factors, including the working delay and the height of the center of gravity. These results may help design better controllers to improve UX in the future.

Our study opens new directions for future work. The controller designing of ODT is a typical human-computer interaction problem. Due to the randomness of human motion and ODT directly affects human walking, the research in this area will involve the estimation of human motion and complex control theory. We plan to introduce the height of the center of gravity into the controller designing to achieve a better control effect. Moreover, in order to improve the prediction performance of the controller, the estimation of human motion will be added to the controller as auxiliary information. Furthermore, we will consider applying reinforcement learning strategies to solve such human-computer interaction problems. As for the UX evaluation, in order to be able to distinguish and to improve the UX brought by different controllers, we will build a single quantitative metric that can directly reflect the UX. The metric can be used to objectively guide the design of the controller and help to formulate a reward function in the reinforcement learning strategy.

%% if specified like this the section will be committed in review mode
\acknowledgments{
%The authors wish to thank A, B, and C. This work was supported in part by
%a grant from XYZ (\# 12345-67890).
This work was supported by 
 the National Natural Science Foundation of China (Grant No. 61773118, Grant No. 61973083),
 Science and Technology Project of State Grid Corporation of China (Intelligent operation and maintenance technology of distributed photovoltaic system SGTJDK00DYJS2000148), the Key Laboratory of Measurement and Control of Complex Systems of Engineering, Ministry of Education, Nanjing, 210096, P. R. China, the Open Project Program of State Key Laboratory of
Virtual Reality Technology and Systems, Beihang University (No. VRLAB2021C04)}.
\balance
\bibliographystyle{abbrv-doi}

\bibliography{template}
\end{document}